\documentclass[reprint, superscriptaddress,nofootinbib, amsmath,amssymb,aps, prb, floatfix]{revtex4-2}

\usepackage{xcolor}

\usepackage{bm}
\usepackage{ulem}
\usepackage{mathtools}
\usepackage{lipsum}
\usepackage{amsmath}
\usepackage{amssymb}
\usepackage{amsfonts}
\usepackage[colorlinks=true,linkcolor=black]{hyperref}
\usepackage[noabbrev]{cleveref}
\usepackage{orcidlink}

\definecolor{dkblue}{rgb}{0,0,0.5}
 \newcommand{\expct}[1]{\left\langle #1 \right\rangle}

\usepackage{amssymb}
\usepackage{amsmath}
\usepackage{amsfonts}
\usepackage{braket}
\usepackage{graphicx}
\usepackage{epstopdf}
\usepackage{epsfig}
\usepackage[toc,page,title,titletoc,header]{appendix}
\usepackage{dsfont,amsthm,amsbsy}
\usepackage{hyperref}
\usepackage[xspace]{ellipsis}
\usepackage{cancel}
\usepackage{tikz}
\usetikzlibrary{automata, positioning}

\begin{document}
\title{Interacting electrons in silicon quantum interconnects}

\author{Anantha~S.~Rao\orcidlink{0000-0001-6272-0327}}
\email{anantha@umd.edu}
\affiliation{Department of Physics, University of Maryland, College Park, Maryland 20742, USA}
\affiliation{Joint Center for Quantum Information and Computer Science, National Institute of Standards and Technology and University of Maryland, College Park, Maryland 20742, USA}

\author{Christopher~David~White\orcidlink{}}
 \affiliation{Center for Computational Materials Science, U.S. Naval Research Laboratory,  Washington, D.C. 20375, USA}

\author{Sean~R.~Muleady\orcidlink{0000-0002-5005-3763}}
\affiliation{Joint Center for Quantum Information and Computer Science, National Institute of Standards and Technology and University of Maryland, College Park, Maryland 20742, USA}
\affiliation{Joint Quantum Institute, National Institute of Standards and Technology and University of Maryland, College Park, Maryland 20742, USA}

\author{Anthony~Sigillito\orcidlink{0000-0002-4765-9414}}
\affiliation{Department of Electrical and Systems Engineering, University of Pennsylvania, Philadelphia, Pennsylvania 19104, USA}

\author{Michael~J.~Gullans\orcidlink{0000-0003-3974-2987}}
\email{mgullans@umd.edu}
\affiliation{Department of Physics, University of Maryland, College Park, Maryland 20742, USA}
\affiliation{Joint Center for Quantum Information and Computer Science, NIST/University of Maryland, College Park, Maryland 20742, USA}
\affiliation{National Institute of Standards and Technology, Gaithersburg, MD 20899, USA.}

\begin{abstract}
Coherent interconnects between gate-defined silicon quantum processing units are essential for scalable quantum computation and long-range entanglement. 
We argue that one-dimensional electron channels formed in the silicon quantum well of a Si/SiGe heterostructure exhibit strong Coulomb interactions and realize strongly interacting Luttinger liquid physics. 
At low electron densities, the system enters a Wigner regime characterized by dominant $4k_F$ correlations;
increasing the electron density leads to a crossover from the Wigner regime to a Friedel regime with dominant $2k_F$ correlations.
We support these results through large-scale density matrix renormalization group (DMRG) simulations of the interacting ground state under both screened and unscreened Coulomb potentials. 
We propose experimental signatures of the Wigner-Friedel crossover via charge transport and charge sensing in both zero- and high-magnetic field limits.
We also analyze the impact of short-range correlated disorder — including random alloy fluctuations and valley splitting variations — and identify that the Wigner-Friedel crossover remains robust until disorder levels of about 400 \textmu eV. 
Finally, we show that the Wigner regime enables long-range capacitive coupling between quantum dots across the interconnect, suggesting a route to create long-range entanglement between solid-state qubits.
Our results position silicon interconnects as a platform for studying Luttinger liquid physics and for enabling architectures supporting nonlocal quantum error correction and quantum simulation.

\end{abstract}

\maketitle

\section{Introduction}
\label{sec:intro}

\begin{figure}[!ht]
  \includegraphics[width=\columnwidth]{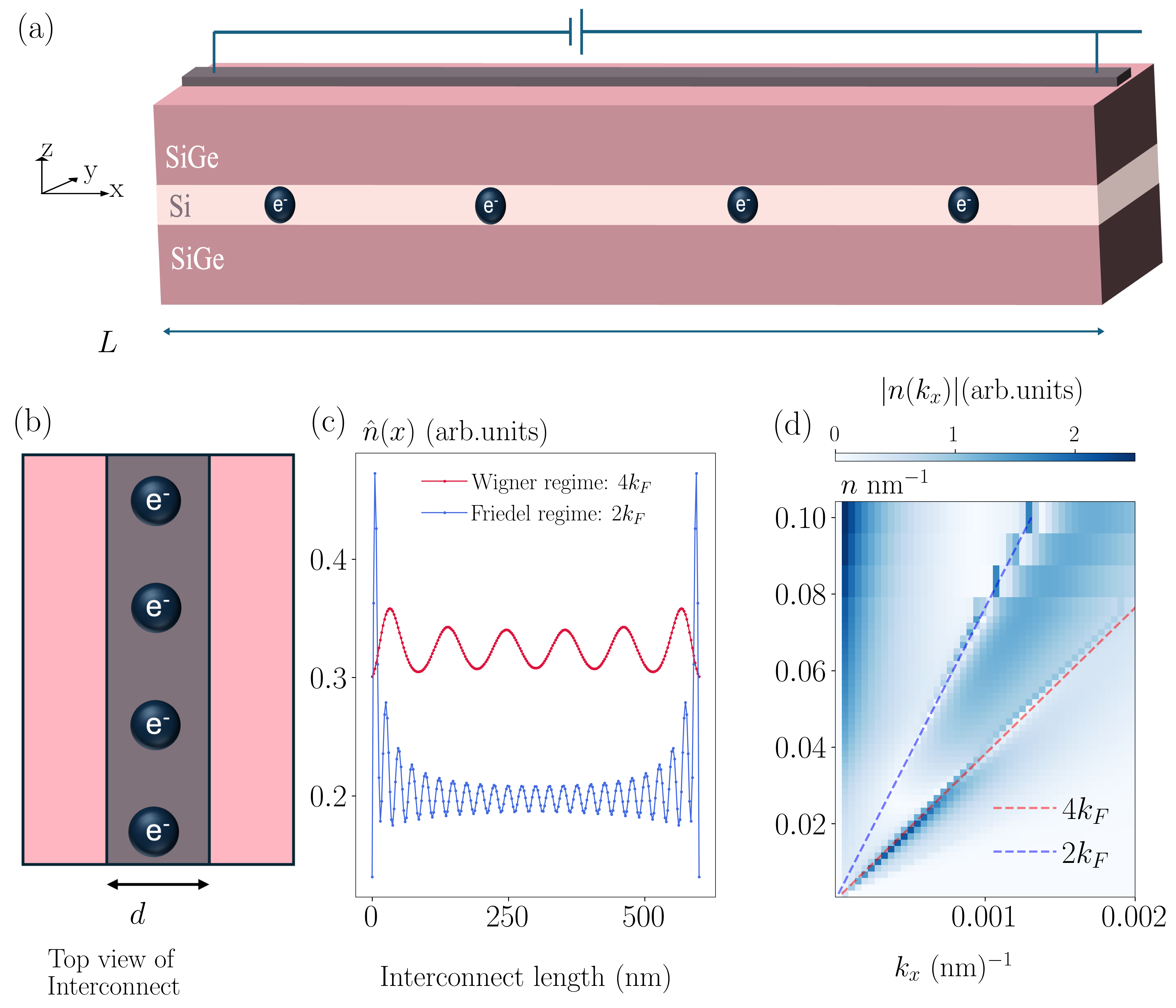}
  \caption{
  (a) Cross-section of the resistive interconnect, consisting of a Si/SiGe heterostructure. A positively biased wire, or ``topgate'', creates a one-dimensional electron channel (1DEC) in the silicon well. (b) Top-view of the interconnect where the topgate creates a 1DEC with a lateral width $d$ under it in the Si well and assists in momentum-incoherent shuttling of electrons between quantum dots. (c) Ground state charge density profile of the 1DEC obtained from density matrix renormalization group calculations in the Wigner (magenta, $n=0.01\text{ nm}^{-1}$) and Friedel (blue, $n=0.083$ nm$^{-1}$) regimes when the external magnetic field is set to zero. The Wigner regime charge density has been offset by 0.3 nm$^{-1}$ for better visualization. (d) The absolute value of the Fourier transform of the ground state charge density within the 1DEC displays a crossover from a potential energy dominated Wigner regime with $4k_F$ correlations to a kinetic energy dominated Friedel regime with $2k_F$ correlations with increasing density, $n \geq 0.05 \text{ nm}^{-1}$ in the interconnect.}
  \label{fig:heterostructure}
\end{figure}

Qubits formed within gate-defined semiconductor quantum dots have demonstrated single- and two-qubit gate fidelities higher than the error-correction threshold and are strong candidates for fault-tolerant quantum computation~\cite{zajac2018resonantly, xue2022quantum, wang2024operating, Mills2022, BurkardEtAl2023, Steinacker2025}.
Scalable devices, however, require \textit{interconnects}~\cite{vandersypen2017interfacing, taylor2005fault, AwschalomPRX2021} to coherently transmit quantum information between distant intermediate-scale qubit registers, overcoming limitations of crosstalk due to the tight gate pitch and high classical wiring density.

Promising approaches to interconnect design include transporting carriers adiabatically or via Landau-Zener transitions across a sequence of quantum dots (spin CTAP, bucket-brigade shuttling)~\cite{gullans2020coherent, mills2019shuttling, Ginzel2020spin, fujita2017coherent, Zwerver2023PRX}; transferring a single carrier through a moving quantum dot formed by clavier gates (conveyor-mode shuttling)~\cite{Seidler2022, Xue2024, Struck2024, Smet2024}; or coupling carriers to microwave photons~\cite{Viennot2015, Samkharadze2018, borjans2020resonant}.
Recently, it was proposed to shuttle carriers through a channel regulated by a single resistive top-gate~\cite{white2024electricalinterconnectssiliconspin}, consisting of a Si-SiGe heterostructure with a positively biased resistive wire on top of the heterostructure.
This nanowire creates a one-dimensional electron channel (1DEC) in the silicon well, and its tunable electric field allows momentum-incoherent transport of electrons from source to target quantum dot.
Although designed for quantum computation~\cite{white2024electricalinterconnectssiliconspin}, 
this electric interconnect also offers a natural platform for exploring the mesoscopic physics of interacting electrons in one dimension, whose study in turn can inform improved device characterization and design. 

Interacting electrons in one dimension form Luttinger liquids that display spin-charge separation and power-law correlations~\cite{tomonagaRemarksBlochsMethod1950,luttingerExactlySolubleModel1963,bouchouleGlanceLuttingerLiquid2025, Fiete2007colloquium},
and exhibit a crossover between distinct low-density interaction-dominated and high-density kinetic-energy-dominated regimes ~\cite{ZianiEtAl2021a,Fiete2007colloquium}. At sufficiently low densities ($na_B \ll 1$, for $n$ the electron number density and $a_B$ the effective Bohr radius) and with unscreened long-range interactions, electrons in one dimension can form a Wigner crystal: a periodic array of widely separated electrons~\cite{wignerInteractionElectronsMetals1934,schulzWignerCrystalOne1993,meyer2008wigner,deshpandeElectronLiquidsSolids2010}. 
In this state, correlations decay more slowly than any power law, and the electron separation is much larger than the electron wavefunction width.
Increasing temperature or density, or introducing sufficient screening, leads to a melting of this Wigner crystal, and can be understood in terms of a quantum analogue of the Lindemann criterion \cite{vuThermalMeltingQuantum2022}. 
At low density and in the presence of screening, electrons are still well separated by Coulomb repulsion and the exchange-induced spin-spin interaction is small, while the charge density exhibits characteristic oscillations at wavevector $4k_F$ (wavelength $1/n$).
We refer to this as the \textit{Wigner regime}.
For higher densities, the kinetic energy dominates and the electrons are separated by Pauli repulsion. 
The characteristic $4k_F$ charge density oscillations of the Wigner regime give way to $2k_F$ Friedel oscillations (wavelength $2/n$); we refer to this as the \textit{Friedel regime}.

Signatures of the Wigner-to-Friedel crossover have been observed in multiple experimental platforms, and are known to persist even with short-range interactions~\cite{Soffing2009Wigner} 
In particular, the electronic compressibility and spatial periodicity of the Wigner regime have been studied in semiconductor quantum wires~\cite{yamamotoNegativeCoulombDrag2006},
in carbon nanotubes~\cite{deshpandeOnedimensionalWignerCrystal2008,lotfizadehBandGapDependentElectronicCompressibility2019a,shapirImagingElectronicWigner2019}, and layer-stacking domain walls of two-dimensional van der Waals heterostructures~\cite{li2024imaging}. 
In the context of semiconductor quantum dots, Wigner molecules have been predicted in parabolic two-dimensional and one-dimensional elongated quantum dots~\cite{Egger1999, Goldberg2024}. 
These Wigner molecules have been predicted to form zigzag one-dimensional chains as the transverse confinement is lowered, as expected for quasi-one-dimensional systems~\cite{hasseStructureCylindricallyConfined1990,piacenteGenericPropertiesQuasionedimensional2004,Meyer2009,Piacente2010,Ho2018PRL,Goldberg2024}.
Elongated quantum dots have also been explored as platforms for quantum dot coupling but are restricted to house only a few electrons and mediate quantum dots through exchange-coupling~\cite{wang2023jellybean, Fedele2021}.
In contrast, a resistive interconnect has very strong transverse confinement with electron motion along the interconnect axis regulated by a single control parameter.

In this work, we investigate the suitability of gate-defined electrical interconnects for probing the crossover from the interaction-dominated Wigner regime to the kinetic energy dominated Friedel regime of a Luttinger liquid.
We perform analytical and numerical studies to characterize the behavior of this system, incorporating relevant features such as interaction screening and disorder relevant to the system.
This complements prior studies on the effects of unscreened Coulomb interactions in one-dimension and its resulting Wigner crystal behavior~\cite{Fano1999, Chou2020}, as well as additional numerical work focusing on the effects of temperature, disorder or device-specific background potential on the spatial charge density of the Wigner crystal~\cite{Fano1999, Piacente2004, VuDasSarma2020}. 
We study the compressibility and density correlations of this system for a range of screening lengths, finding signatures of this crossover as long as the electron spins remain unpolarized.
We also propose transport and charge-sensor signatures of the crossover,
and discuss their robustness against disorder. 
We find that electronic compressibility measurements inform the crossover from the Wigner to the Friedel regime.
Charge sensor measurements help determine the exact crossover density and measure the effects of disorder seen in similar Si/SiGe quantum dot devices.

Crucially, our characterization of the Wigner-Friedel crossover addresses a central challenge in scalable quantum computing: the realization of long-range interconnects between distant quantum modules. 
We find that the strongly interacting nature of the Wigner regime facilitates long-range capacitive coupling that is significantly more effective than in the kinetic-dominated Friedel regime.
This `stiffness' of the Wigner regime allows local chemical potential shifts to propagate across the channel to neighboring quantum dots, establishing a mechanism for information transfer that is robust against realistic disorder.
By identifying the physical regimes where these correlations overcome screening and disorder, our work establishes the gate-defined Wigner channel as a viable, tunable interconnect, paving the way for modular architectures in semiconductor-based quantum computing.

The paper is organized as follows.
In Sec.~\ref{sec:model} we describe our model (electrons in effective mass theory, with unscreened and screened Coulomb interactions),
and in Sec.~\ref{sec:methods} we describe our methods (bosonization and the density matrix renormalization group, or DMRG).
In Sec.~\ref{sec:clean-channel}, we discuss the Wigner-Friedel crossover in the channel in the absence of disorder; and propose two experimental probes---electronic compressibility and charge sensing by a nearby dot.
In addition, we show that results from bosonization qualitatively agree with numerical results.
In Sec.~\ref{sec:results-disordered} we briefly discuss the Wigner-to-Friedel crossover and our charge sensing probe in the presence of short-range correlated disorder.
In Sec.~\ref{sec:cap-coupling-IC} we discuss the performance of the Luttinger liquid as a tunable capacitive coupler between double quantum dots on either side of the channel and effects of short-range correlated disorder.
Finally in Sec.~\ref{sec:summary} we summarize our results and discuss potential future work.

\section{Model and methods}

\subsection{Model}
\label{sec:model}

We consider a resistive interconnect, composed of a silicon heterostructure (Fig.~\ref{fig:heterostructure})---%
a thin layer of silicon between thick layers of silicon-germanium alloy---%
with additional layers and gates on top.
Conduction band bending at the Si-SiGe interface traps electrons along the $z$-axis, leading to the formation of a high mobility two-dimensional electron gas (2DEG) in the silicon layer ($x-y$ plane in Fig.~\ref{fig:heterostructure}(A)). 
A one-dimensional wire on top of the heterostructure is gated to produce a narrow potential energy minimum in the well, confining the electron to a quasi-one-dimensional channel of width $d \approx 10\text{ nm} \text{ -- } 100$ nm.
The energy of the transverse orbital excitations is $E = \pi^2\hbar^2/(2m_td^2)\approx (0.2 \text{ to }20)$ meV where $m_t = 0.19 m_e$ is the transverse effective mass of conduction electrons in Si~\cite{kittel2018introduction}.
We take the electron Hamiltonian to be purely one dimensional
and take the transverse positional degree of freedom into account via a UV regularization of the interaction. 

\subsubsection{Clean system}

The dynamics of $N$ interacting electrons in a one-dimensional quantum channel are given by the effective mass Hamiltonian
\begin{align}\label{eq:ham-em}
\begin{split}
    \hat{H} &= \hat{H}_0 + \hat{V}_{ee} \\
    \hat{H}_0 &= \sum_{\sigma} \int  dx\; \hat{\psi}^{\dagger}_{\sigma}(x)\left[\frac{-\hbar^2}{2m_t}\frac{\partial^2}{\partial x^2} + V_{\mathrm{top}}(x) \right] \hat{\psi}_{\sigma}(x) \\
    \hat{V}_{ee} &= \frac{1}{2}\sum_{\sigma, \sigma'}\int dx\; dx'\;  U(x-x')\\
    &\qquad\qquad\quad\times \hat{\psi}_{\sigma}^{\dagger}(x) \hat{\psi}_{\sigma'}^{\dagger}(x')\hat{\psi}_{\sigma'}(x')\hat{\psi}_{\sigma}(x)
\end{split}
\end{align}
where $\hat{\psi}_\sigma(x)$/$\hat{\psi}_\sigma^\dagger(x)$ are fermionic annihilation and creation operators and  $\sigma\in\{\uparrow,\downarrow\}$ is the spin index.
Here $x$ denotes the position along the length of the interconnect,  $V_{\text{top}}(x)$ is the electrostatic potential experienced by each electron due to the top-gate and device confinement, and $U(x-x')$ is the electron-electron interaction potential.
We take $V_{top}$ to be an infinite box-potential that restricts electrons within the 1D channel. 

The electrons interact via a screened Coulomb potential
\begin{align}
    \label{eqn:screened-interaction}
    U(r) = \frac{e^2}{4\pi\epsilon_r\epsilon_0 d}\frac{e^{-r/D}}{\sqrt{1 + r^2/d^2}}
\end{align}
where $r\equiv |x-x'|$ is the longitudinal separation,
$D$ is an effective screening length set by device geometry (in particular, nearby metallic gates) \cite{Davis1973, Knorzer2022-LRE},
and $d$ is an effective transverse length scale that characterizes the width of the interconnect, while $\epsilon_r$
and $\epsilon_0$ are the relative permittivity for Si and vacuum permittivity, respectively.

The channel width $d$ regularizes the UV divergence in the Coulomb interaction~\cite{goldAnalyticalResultsSemiconductor1990, schulzWignerCrystalOne1993},
and at the same time accounts for the nonzero channel width.
If two electrons are close in the $x$ direction, along the channel, we expect
them to be a distance $\approx d$ apart in the $y$ direction (in-plane, transverse to the channel).
As the density increases this effect leads to a zigzag regime, in which the electrons alternate between the two edges of the channel, and a richer phase diagram~\cite{hasseStructureCylindricallyConfined1990,piacenteGenericPropertiesQuasionedimensional2004,Meyer2009,Piacente2010,Goldberg2024}.
To treat this physics in detail, one would include transverse orbital excited states in the single-electron basis.
For low densities $n \lesssim d^{-1} \approx 0.1\text{ nm}^{-1}$, however,
the regularization $r \to \sqrt{r^2 + d^2}$ takes this effect into account.

We discretize the continuum effective mass model \eqref{eq:ham-em} to obtain a tight-binding representation:
\begin{align}
  \begin{split}\label{eqn:lattice-hamiltonian}
    \hat{H} &= -t \sum_{j=1, \sigma}^{N_{tot}-1} \left(\hat{c}_{j,\sigma}^{\dagger} \hat{c}_{j+1,\sigma} + h.c.\right) \\
      &\qquad + \frac 1 2 \sum_{i, j,\sigma,\sigma'} U(x_i-x_j) \hat{n}_{i,\sigma} \hat{n}_{j,\sigma'}\;,
  \end{split}
\end{align}
where $N_{tot}$ is the number of lattice sites, $\hat{c}_{j,\sigma}$ $(\hat{c}_{j,\sigma}^\dagger)$ annihilates (creates) a spin-$\sigma$ electron on lattice site $j$, $\hat{n}_{i,\sigma}=\hat{c}_{i,\sigma}^\dagger\hat{c}_{i,\sigma}$ is the corresponding fermionic number operator, and the hopping amplitude is $t=\hbar^2/(2m_t a^2)$. The lattice positions are denoted $x_j = j a$ for lattice cutoff $a$, and we have $N_{tot} = L/a$ for channel length $L$.

When $a$ is smaller than all the length scales of interest,
the lattice model \eqref{eqn:lattice-hamiltonian} reduces to the continuum effective mass Hamiltonian \eqref{eq:ham-em}.
In particular, for lattice model filling $nL/N_{tot} \ll 1/2$,
i.e. typical electron spacing $n^{-1} \gg a$, the hopping in \eqref{eqn:lattice-hamiltonian} reduces to the quadratic kinetic energy in $\eqref{eq:ham-em}$,
and for $a \ll d,D$ the discretization appropriately captures the screened interaction.
We take $a = 2.5$ nm, and check for convergence in this lattice cutoff in Appendix.~\ref{sec:device-parameters}.

\subsubsection{Disorder model}
\label{sec:disorder-model}

In addition to the clean system, we consider two kinds of disorder effects: a random local potential, which couples only to the electron density, and valley disorder, which couples to an additional two-dimensional valley degree of freedom.
The random local potential is
\begin{align}
  \hat{H}_{\text{disorder}} &= \sum_{i=1}^{N_{tot}} W(x_i) \hat{n}_i
\end{align}
where $W$ are Gaussian random variables with
\begin{align}
    \begin{split}
    \expct{ W(x) } &= 0 \\
    \expct{ W(x) W(x') } &\simeq W^2\,f_{\xi}(x-x');
    \label{eqn:short-range-disorder}
  \end{split}
\end{align}
here $\langle \cdot \rangle$ is an average over disorder realizations, $W^2$ sets the disorder variance and $f_{\xi}(r)$ is a short-range correlation function. 
We use a white noise idealization $f_\xi(r)=\delta_{r,0}$ for the rest of our analysis.
The standard deviation $W$ therefore encodes the effective strength of short-range electrostatic disorder after averaging over the lattice cutoff and transverse profile.
Possible sources of disorder include
random fluctuations in the concentrations of silicon and germanium at the Si/SiGe interface (`alloy disorder');
monolayer steps or interface roughness at the Si/SiGe interface, and
charged defects near the capping layer at the surface of the device.
Alloy disorder has correlation lengths of a few  nanometers or shorter~\cite{Marcks2025-VSC},
which can be treated in effective mass theory as a delta function.
Monolayer steps and charged defects at the heterointerface have longer correlation lengths, but will have short-correlation length components; 
we leave the treatment of long-range-correlated disorder to future work.

In silicon, the valley degree of freedom provides an additional two-level subspace for each electron~\cite{BurkardEtAl2023}. 
The valley splitting is the energy difference between the two lowest lying valley states that the electron could occupy, and is strongly influenced by the relative concentration profiles of silicon and germanium along the z axis of the device~\cite{losert2023practical}. 
Furthermore, the valley degree of freedom can substitute for spin in the context of the $4 k_F \to 2 k_F$ crossover.
For the purposes of this work, we ignore spin-valley coupling in silicon.
This simplifies the analysis by avoiding the rich $8 k_F \to 4 k_F \to 2 k_F$ landscape that arises from the interplay of the spin, valley and charge sectors.
Instead, we consider either spinful (or spinless) electrons with no valley degree of freedom, as above (that is, we assume the valley splitting exceeds other energy scales),
or we consider electrons with no spin degree of freedom, as here (that is, we assume the Zeeman splitting exceeds other energy scales).
Short-range correlated disorder like fluctuations in alloy concentration also affects valley splitting, leading to valley disorder. 
We analyze the effect of the valley degree of freedom on the interconnect by considering an additional term to the channel Hamiltonian \eqref{eqn:lattice-hamiltonian} given by
\begin{align}
    \label{eqn:valley-ham-2}
    \hat{H}_{valley} = \sum_{i=1}^{N_{tot}} | \Delta_{v}(x_i)| \left(\cos{\phi(x_i)}\hat{\tau}^{x}_i + \sin{\phi(x_i)}\hat{\tau}^{y}_i \right)
\end{align}
where $\hat{\boldsymbol\tau}_i$ are Pauli matrices acting in the two-dimensional valley subspace at site $i$ (i.e. the effective pseudo-spin basis states for the two lowest valley minima), $2\Delta_v(x_i)$ is the local energy splitting between the valley states, and the phase $\phi(x_i)$ encodes the relative mixing of the valley basis states (determined microscopically by the local atomic-scale interface profile and step disorder).

We assume a strong external magnetic field such that the spin degree of freedom is fully polarized; thus, the spin index is omitted, and the system effectively behaves as spinless fermions with a valley pseudo-spin index.
The first term in \eqref{eqn:lattice-hamiltonian} represents kinetic hopping, which we model as purely intra-valley (conserving the valley index during tunneling).
The second term describes the long-range Coulomb repulsion.
The term $\hat{H}_{\text{valley}}$ captures any inter-valley coupling.
We sample each onsite $\Delta_v$ from a complex normal distribution with mean $\bar{|\Delta_v}| = 9.58$ \textmu eV and standard deviation $1.03$ meV (See Appendix.\ref{appendix:valley-variance} for details).
$2|\Delta_v(x_i)|$ is the local energy splitting between the valley states at site index $i$, and the phase $\phi(x_i) = \arg |\Delta_v(x_i)|$ encodes the relative mixing of the valley basis states. 
In this basis, disorder manifests as a local rotation of the valley eigenstates, creating an effective inter-valley coupling that varies spatially. 
Spatial variations of $\Delta_v$ or abrupt phase changes can cause non-adiabatic transitions between the instantaneous valley eigenstates (valley mixing), effectively scattering the electron from the ground valley state $v_-$ into the excited valley state $v_+$; we therefore study valley disorder alongside electrostatic disorder in Sec.~\ref{sec:results-disordered}.
We take the mean valley splitting disorder $\bar{|\Delta_v}|  > 0$ in accordance with experiment and prior theoretical work.
In principle this could lead to an additional crossover: for sufficiently low density the nonzero average valley splitting could lead to valley polarization.
We expect, however, that that effect is negligible compared to any of valley splitting disorder, the kinetic energy, or the interaction energy.

\begin{figure*}
    \centering
    \includegraphics[width=\linewidth]{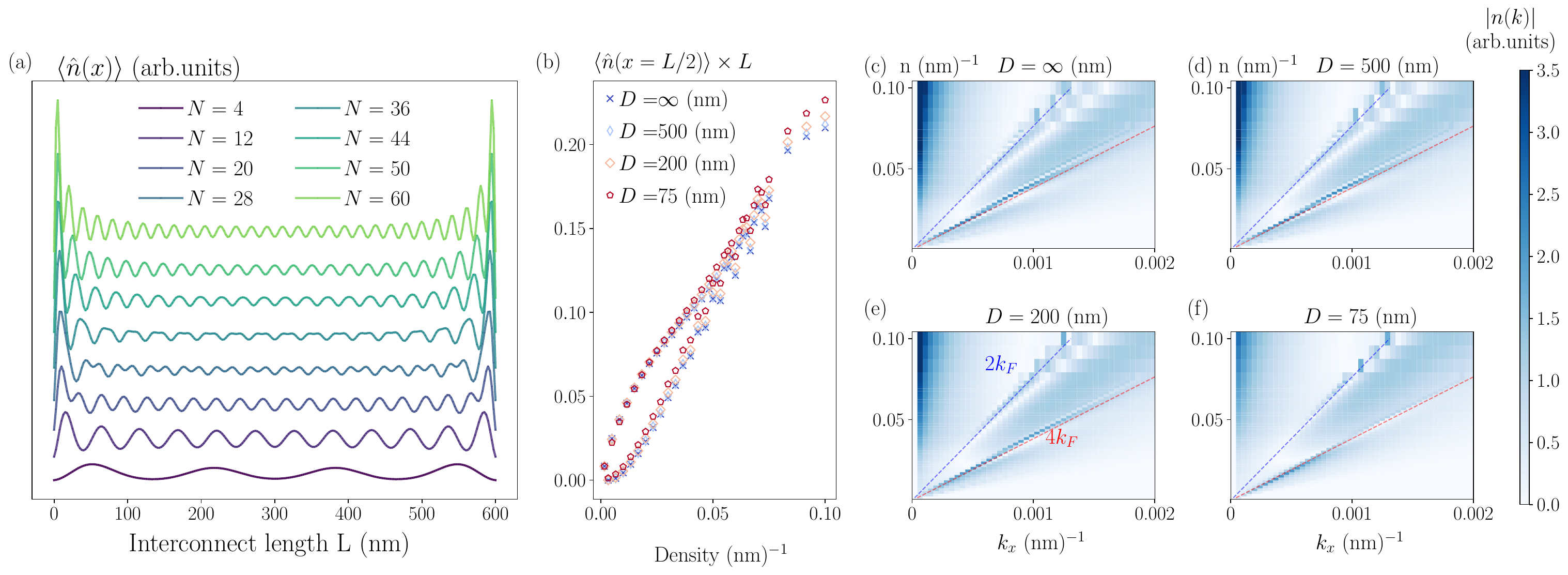}
    \caption{Wigner crystal regime as the ground state of electrons within the one-dimensional Si-well. (a) Total charge density of the ground state in a $L=$600 nm interconnect with different numbers of electrons interacting via the screened-Coulomb interaction with screening length $D=75$ nm (charge density for every particle number is offset by an arbitrary constant). 
    As the number of electrons in the channel increases, we observe a crossover from a potential energy dominated Wigner regime with $4k_F$ correlations to kinetic energy dominated Friedel regime with $2k_F$ correlations. In the Wigner regime, $N$ electrons in the channel display $N$ distinct peaks, whereas the Friedel regime only displays $N/2$ such peaks. (b) Charge density at the midpoint of the interconnect for different screening lengths. For low densities, we observe strong oscillations in the charge density with the parity of the electron number, which vanishes in the higher-density Friedel regime; this observation informs our later charge-sensing protocol. (c-f) Absolute value of the Fourier transform of charge density plotted as a function of the wavevector and electron density for different screening lengths. Red (blue) lines denote the $4k_F(2k_F)$ wavevectors $(k_F=n\pi/2)$ for each electron density.
    As the density of the electrons increases beyond 0.05 nm$^{-1}$ or around $(na_B)^{-1} \approx 6$ in the channel, there is a crossover in the peak of the Fourier transform of the ground state wavefunction, which is observed across different effective screening lengths.
    }
    \label{fig:spinful-screening}
\end{figure*}

\subsection{Methods} \label{sec:methods}
We obtain the ground state properties of the interacting one-dimensional electron system analytically, using bosonization~\cite{schulzWignerCrystalOne1993, giamarchi2003quantum, VoitPRB1993}, and numerically, using the density matrix renormalization group (DMRG) \cite{White1992, schollwoeckDensitymatrixRenormalizationGroup2011}. 
Bosonization provides analytical insights into the Luttinger liquid parameters in the thermodynamic limit; DMRG calculations give energy levels, compressibilities, capacitive couplings, and finite-size and disorder effects.

\subsubsection{Bosonization}
\label{ssec:bosonization}

One-dimensional interacting fermions can be solved by bosonization~\cite{schulzWignerCrystalOne1993, giamarchi2003quantum, VoitPRB1993} where spin-charge separation enables separate treatment of the spin and charge degrees of freedom as independent bosonic modes. 
The Hamiltonian \eqref{eq:ham-em} can be recast as the continuum model
\begin{subequations}
\begin{align}
    H &= H_c + H_s\\
    H_{c/s} &= \frac{u_{c/s}}{2\pi}\int dx \left[ K_{c/s}(\partial_x\theta_{c/s})^2 + K_{c/s}^{-1}(\partial_x \phi_{c/s})^2\right]
\end{align}
\end{subequations}
where $H_{c/s}$ is the charge/spin-sector Hamiltonian;
$\{\theta, \phi \}_{c/s}$ are the bosonic charge/spin fields and their conjugates;
$u_{c/s}$ and $K_{c/s}$ are the corresponding velocity and Luttinger parameter, respectively, which depend on the Fourier transform of the interaction potential $\tilde{U}(q)$;
and $v_F=k_F/m$ is the Fermi velocity. 
The charge-sector Luttinger parameter $K_c$ is a measure of the strength of interactions in the system; $K_c \ll 1$ for strong repulsive interactions; $K_c = 1$ for non-interacting, free fermions; and $K_c > 1$ for attractive interactions.
At low temperatures and ignoring Umklapp scattering, interactions of the form \eqref{eqn:screened-interaction} give Luttinger parameter $K_c$ and velocity $u_c$ \cite{giamarchi2003quantum}
\begin{align}
\begin{split}
    K_c &= \sqrt{\frac{\pi v_F + g_{4}/2 - g_{2}/2}{\pi v_F + g_{4}/2 + g_{2}/2}}, \\
    u_c &= \frac{1}{\pi} \sqrt{\left(\pi v_F + \frac{g_4}{2}\right)^2 - \left(\frac{g_2}{2}\right)^2}
    \end{split}
\end{align}
where the couplings $g_{4}(g_2)$ are the same (opposite) branch scattering coefficients. For large magnetic fields, the spins polarize, resulting in system of spinless fermions evolving according to $H_c$, and the couplings are approximately
\begin{align}
\begin{split}\label{eqn:g2g4}
    g_4 &\simeq \tilde{U}(0), \quad g_2 \simeq \tilde{U}(0) - \tilde{U}(2k_F).
\end{split}
\end{align}
We compute these numerically for our interaction potential, and also provide approximate analytical expressions for $\tilde{U}(0)$ and $\tilde{U}(2k_F)$ in Appendix~\ref{appendix:bosonization}. 
We note that the bosonization approximation relies on a linearization about the two Fermi points corresponding to left and right moving excitations; for sufficiently low densities, where the relevant Fermi points lie close to the bottom of the $\approx k^2$ potential, we expect this assumption to break down.

\subsubsection{Density Matrix Renormalization Group}

DMRG \cite{White1992, schollwoeckDensitymatrixRenormalizationGroup2011} is based on the matrix product state (MPS) representation of many-body wavefunctions. 
An MPS expresses the quantum state of a chain of sites in terms of products of local tensors whose dimension is controlled by a bond dimension, $\chi$.
This provides a systematic way to approximate states with limited entanglement, such as ground states of one-dimensional Hamiltonians.
In practice, the bond dimension determines the accuracy with which correlations and entanglement can be represented.

The DMRG algorithm variationally optimizes the MPS by minimizing the expectation value of the Hamiltonian through an iterative sweeping procedure: local tensors are updated site by site while keeping the rest of the system fixed, and the effective environment is built from the surrounding tensors.
Our threshold for convergence is achieved when repeated sweeps result in a ground state energy difference of $< 10^{-8} \times t$.
Our Hamiltonian includes long-range Coulomb interactions, which are represented in DMRG using a matrix product operator (MPO).
We perform these calculations with the ITensor library \cite{fishmanITensorSoftwareLibrary2022}, which provides flexible and efficient implementations of both DMRG and MPO constructions.
This framework enables us to explore how the ground state charge density and correlations evolve with electron density, screening length, and disorder.

\section{Results for the Clean channel}
\label{sec:clean-channel}

\subsection{Wigner-to-Friedel crossover: phenomenology}
\label{sec:wigner-friedel-crossover-results}

We obtain the ground state of the many-body Hamiltonian
\eqref{eqn:lattice-hamiltonian} with $N$
electrons through density-matrix renormalization group (DMRG) calculations. 
First, we study the charge-sector properties of the ground state when the magnetic field through the interconnect is set to zero ($B=0$), for different effective screening lengths.
In the low density limit, the ground state charge density shows $N$ peaks (Fig.~\ref{fig:spinful-screening}a),
the mid-channel density displays a clean even-odd alternation (Fig.~\ref{fig:spinful-screening}b), and the Fourier transform of the density is dominated by a peak corresponding to the Wigner-regime $4 k_F$.
As electrons are added to the channel, at a certain density ($n \approx 0.05 \text{ nm}^{-1}$),
the number of peaks in the charge density becomes $N/2$ (Fig.~\ref{fig:spinful-screening}a).
At approximately the same density, the even-odd alternation in the midpoint density ceases and the Friedel-regime $2k_F$ peak in the density Fourier transform becomes the dominant peak, replacing the Wigner-regime $4k_F$ peak.
We observe the same behavior across screening lengths.

We can characterize the crossover density by the Wigner ratio $r_s = r/a_B$, where $r = 1/n$ is the average electron spacing and $a_B=3.27$ nm is the effective Bohr radius in silicon.
The crossover density $n\approx 0.05\text{ nm}^{-1}$ corresponds to $r_s \geq 6$,
which is lower than the corresponding estimate in two-dimensions~\cite{TanatarPRB1989, FalakshahiPRL2005} and similar to the results obtained in ~\cite{li2024imaging, shapirImagingElectronicWigner2019}. 

\begin{figure*}[t]
    \centering
    \includegraphics[width=\linewidth]{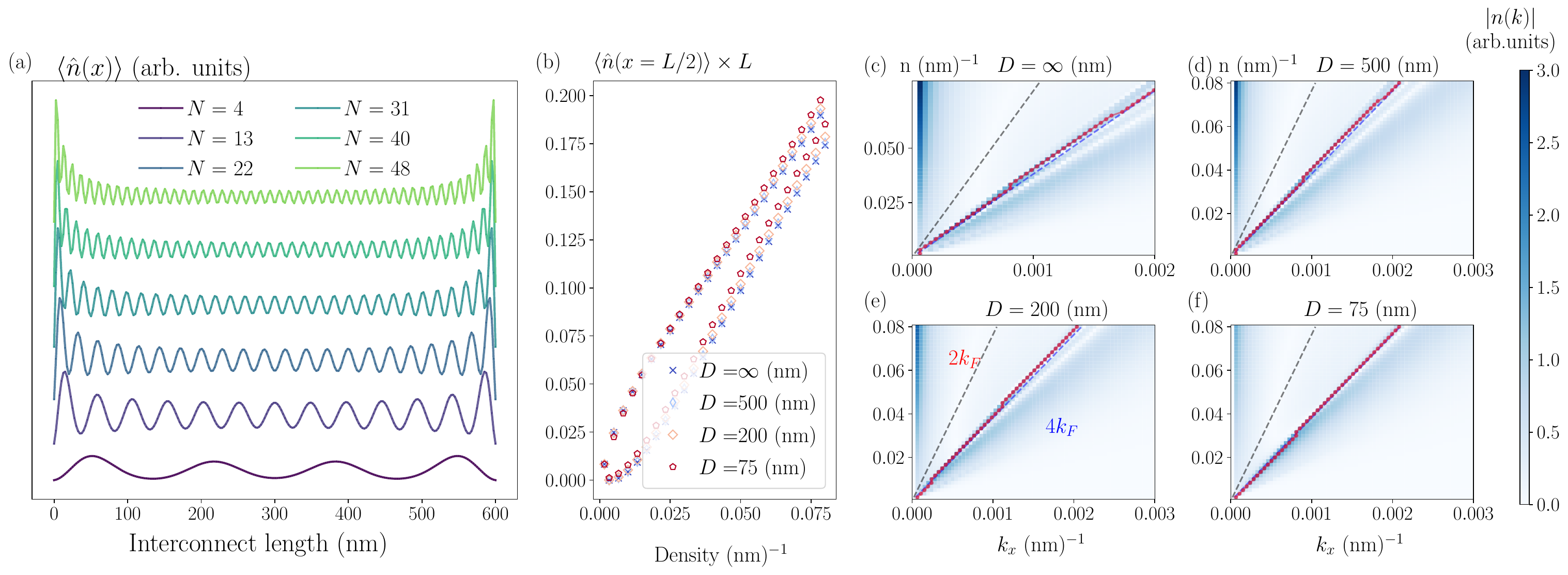}
   \caption{Wigner crystal regime as the ground state of electrons within the one-dimensional Si-well at high-magnetic fields. (a) Charge density of the ground state in a 600nm interconnect with different number of electrons interacting via the screened-Coulomb interaction with screening length $d=75$ nm (charge density for every particle number is offset by a constant). 
    As the number of electrons in the channel increases, we do not observe a crossover from a Wigner-dominated $4k_F$ regime to a Friedel-dominated $2k_F$ oscillation regime. (b) Charge density at the midpoint of the interconnect for different screening lengths. In contrast to Fig.~\ref{fig:spinful-screening}, we observe strong oscillations in the charge-density with odd-even number of electrons up to the highest densities we consider. (c-f) Absolute value of the Fourier transform of charge density plotted as a function of the wavevector and electron density for different screening lengths. (Blue, red, black) lines denote $(2k_F, k_{max}, 4k_F)$ for each electron density. As the density of the electrons increases, the peak of the Fourier transform of the ground state wavefunction remains at $4k_F$ consistently across all screening lengths.}
    \label{fig:clean-system-wc-spinless}
\end{figure*}

The signatures of the Wigner-to-Friedel crossover result from dimerization of spin-up and spin-down electrons.
In a strong magnetic field all spins are polarized
and the electrons can be treated as spinless fermions,
so one does not expect such signatures.
For comparison, we show simulations of the ground state for spinless interacting electrons in our channel in Fig.~\ref{fig:clean-system-wc-spinless}.
Indeed, we see that the Wigner regime's signatures persist to the highest densities we study:
for all those densities
there are $N$ density peaks (Fig.~\ref{fig:clean-system-wc-spinless}a),
the midpoint density displays even-odd alternation (Fig.~\ref{fig:clean-system-wc-spinless}b),
and the Fourier transform of the density is dominated by the $4k_F$ peak.

\subsection{Wigner-to-Friedel crossover: experimental signatures }
\label{sec:experiments}
So far we have probed the density-dependent crossover from the $4k_F$ Wigner-dominated regime to the $2k_F$ Friedel dominated regime within our quasi-one dimensional interconnect. 
But how does one experimentally verify various properties of Wigner regime and observe this crossover?
Based on the ground states obtained through our simulations, we study two main features that may be observed and verified through experiments with the interconnect---strong electronic correlations and periodic spatial dependence. 
We propose two different experiments to probe the electronic compressibility and the charge distribution in the strongly interacting Wigner regime.

\subsubsection{Transport}
\begin{figure*}
    \centering
    \includegraphics[width=\linewidth]{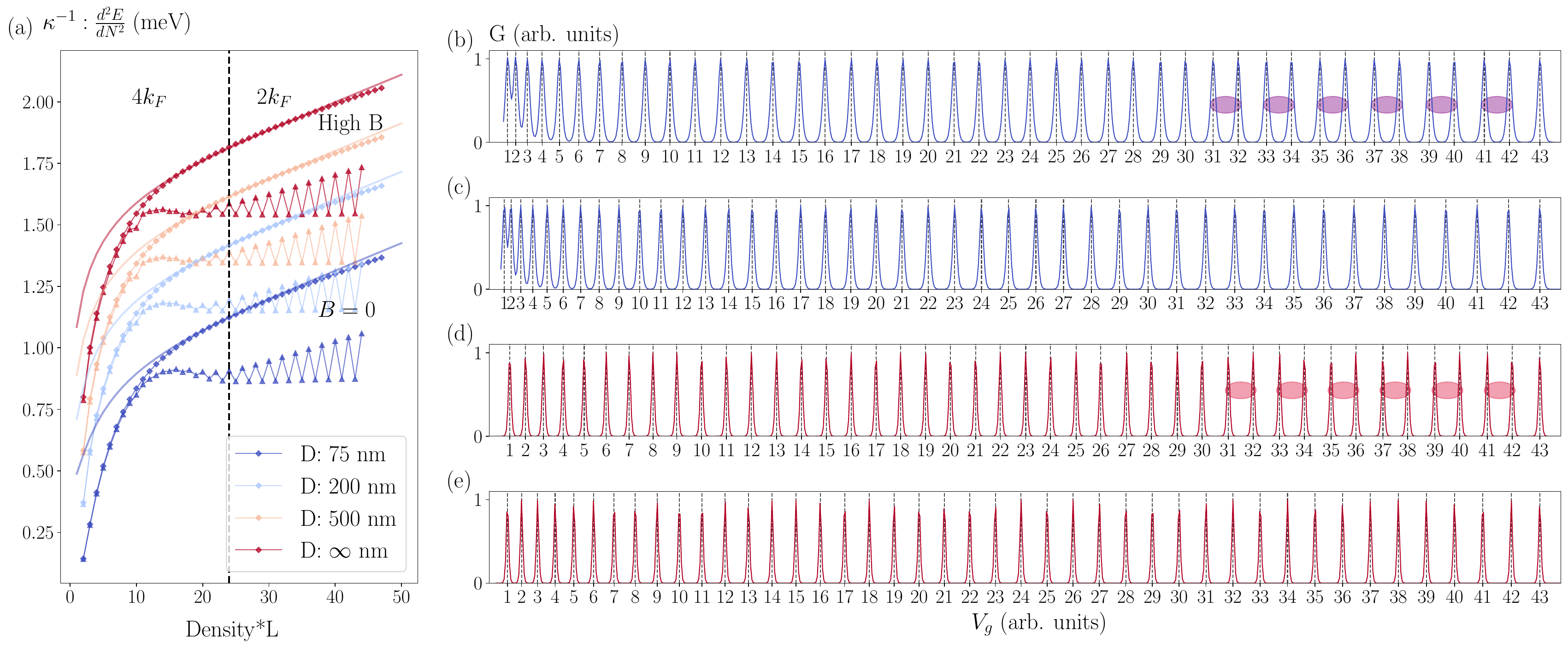}
    \caption{(a) Electronic incompressibility $\kappa^{-1}$ of the interconnect exhibits a strong decrease with decreasing density for both zero and high magnetic field B in the interconnect. 
    $\kappa^{-1}$ in the high magnetic field limit (diamond markers)  shows a monotonic increase with both density and screening lengths. 
    Results from spinless bosonization theory are plotted as solid lines and show excellent agreement with results from DMRG at high densities, while exhibiting similar qualitative trends at lower densities. In the zero B limit (triangular markers), $\kappa^{-1}$  exhibits two distinct regimes---at low densities, there is a monotonic increase and at higher densities, strong oscillations occur with particle number. This is absent in the high B limit where electronic spins are polarized  and $\kappa^{-1}$ increases with additional particles. 
    (b) Phenomenological conductance $G$ of the interconnect with the number of electrons in the channel determined from the incompressibility for $D=75$ nm. Elliptical patches indicate two nearby peaks of opposite spin. At low densities, the spacing between peaks, given by $\kappa^{-1}$ is small, while at high densities, $\kappa^{-1}$ and the gap between peaks is found to alternate as electrons of opposite spin either pair-up or remain unpaired in the Luttinger liquid. (c) In the high $B$ limit, the spacing between conductance peaks continues to monotonically increase with the number of electrons. (d,e) Conductance for the Coulomb interaction without screening $(D=\infty)$ shows alternations in the peak-spacing in the low-B limit and a monotonic increase in the high-B limit.
    }
    
    \label{fig:compressiblity}
\end{figure*}
The relative importance of electron–electron interactions and kinetic energy can be probed directly through transport measurements. 
In particular, we propose studying Coulomb blockade in transport through the interconnect as a way to detect the crossover from the Wigner to the Friedel regime. 
Transport signatures---such as the spacing and structure of conductance peaks---are controlled by the electronic compressibility, $\kappa$~\cite{Lotfizadeh2019PRL, tansElectronElectronCorrelations1998, deshpandeOnedimensionalWignerCrystal2008} defined via
\begin{align}
    \label{eqn:elec_compressibility}
    \kappa^{-1} = \left(\frac{d^2E}{dN^2}\right) = \frac{d\mu}{dN},
\end{align}
where $E$ is the ground state energy of the many-body system, $N$ is the number of electrons, and $\mu = dE/dN$ is the chemical potential of the system. 
Transport experiments through the interconnect can determine the electronic compressibility since conductance occurs only when the interconnect gate voltage $V_G$ aligns the chemical potential of the channel with the bias window between source and drain dot voltages, $V_{SD}$.
In that window, the channel is not under blockade and current can be measured through the channel, with
the spacing of conductance peaks set by $\kappa^{-1}$~\cite{Simmel1999, Berkovits1997}.
The resulting inverse compressibility can then be approximated by
\begin{align}\label{eqn:inv-compressibility-discrete}
    \kappa^{-1} \approx \frac{\Delta \mu(N)}{\Delta N} = \frac{E(N+1) - 2E(N) + E(N-1)}{\Delta N}\;.
\end{align}

Eq.~\ref{eqn:inv-compressibility-discrete} gives the compressibility in experiment and in DMRG.
In bosonization, it can be directly calculated in the thermodynamic limit as
\begin{align}
    \kappa^{-1} = \pi v_F + \frac{1}{2}(g_4 + g_2)
\end{align}
where $v_F$ is the fermi velocity of electrons and $g_4, g_2$ are the scattering coefficients for electrons on the same and opposite branch respectively.
($g_2$ and $g_4$ are given in Eq.~\eqref{eqn:g2g4}).

We evaluate the compressibility from the ground-state energies obtained from DMRG and from bosonization theory.
The evolution of these quantities as a function of density and screening length provides a direct signature of the Wigner–Friedel crossover.
In the regime of strong correlations and low density, such as the Wigner regime, $\kappa^{-1}$ decreases sharply with decreasing density. 
This trend has been observed experimentally in carbon-nanotubes~\cite{Lotfizadeh2019PRL} and is consistent with theoretical predictions~\cite{levitov2003narrow, berkovits1997compressibility}.

We present our results for the inverse compressibility in Fig.~\ref{fig:compressiblity}(a) where we determine $\kappa^{-1}$ from the bosonization expression and from DMRG via finite differences of the ground-state energy as a function of particle number. 
When a net zero magnetic field is maintained in the interconnect, the inverse compressibility exhibits a concave downturn with \textit{decreasing} density in the Wigner regime. 
For $N \gtrsim 24$ electrons, oscillations appear: when the next electron occupies a spin state that leaves an unpaired electron, $\kappa^{-1}$ increases, while pairing of spins lowers $\kappa^{-1}$.
This spin-dependent alternation is a direct signature of the underlying ground-state configuration.
The behavior in a high magnetic field through the interconnect is qualitatively similar but with key differences.
For the first ten electrons, the compressibility is nearly independent of magnetic field, and peak spacings are unchanged. 
At higher densities, however, $\kappa^{-1}$ grows linearly with density, with slope determined by the Fermi velocity, reflecting a progressively more incompressible system. 
Our bosonization calculations capture this trend and agree with results from DMRG at high densities.
Bosonization also qualitatively captures the concave nature of $\kappa^{-1}$ in the Wigner regime, which arises from density-dependent logarithmic contribution to $\tilde{U}(2k_F)$, originating in the $1/r$ nature of the potential at short distances.
We expect deviations at small $n$ to arise partially from the breakdown of the linearized spectrum near the bottom of the quadratic potential.

To model the Coulomb blockade conductance peaks, we approximate the conductance through the channel as
$G = \sum_{x_{peak}} \cosh^{-2}{(\Omega(V_G-V_G^{peak}))}$,
where $V_G^{peak}$ are the peak positions predicted by $\kappa^{-1}$~\cite{Beenakker1991}.
The parameter $\Omega$ controls a phenomenological thermal broadening on top of our $T = 0$ DMRG calculations; $\Omega = \alpha/2k_BT$, where $\alpha$ is the relative lever-arm of the topgate.
To resolve these conductance peaks, the thermal broadening must be smaller than the level spacing, which is $\approx \kappa^{-1},$ the inverse compressibility. Since $\kappa^{-1} \gtrsim 0.25 \text{ meV}$, we thus require $T \lesssim 2$ K.
We take $T = 500$ mK; at $\alpha = 5$ meV/V this gives $\Omega = 46 \text{ V}^{-1}$. 

Our results for the pattern of conductance peaks in the zero- and high-magnetic field limits for both screened and unscreened interactions are shown in Fig.~\ref{fig:compressiblity}(b,d) and Fig.~\ref{fig:compressiblity}(c,e) respectively.
The first peak is placed at an arbitrary gate voltage $V_G = 1$.
We observe that in the high-density, Friedel regime, conductance peaks appear in pairs, with successive, closely spaced peaks corresponding to opposite spins, followed by a larger gap before the next electron enters. 
The overall trend is that shorter screening lengths reduce $\kappa^{-1}$, since electrons can be more easily compressed once placed beyond their screening length, and this results in a reduced spacing between conductance peaks.
Together, these results demonstrate how transport experiments can provide estimates of compressibility, and provide a concrete route to identify the Wigner–Friedel crossover in a finite interconnect.

\subsubsection{Charge sensing}

\begin{figure}
    \centering
    \includegraphics[width=\linewidth]{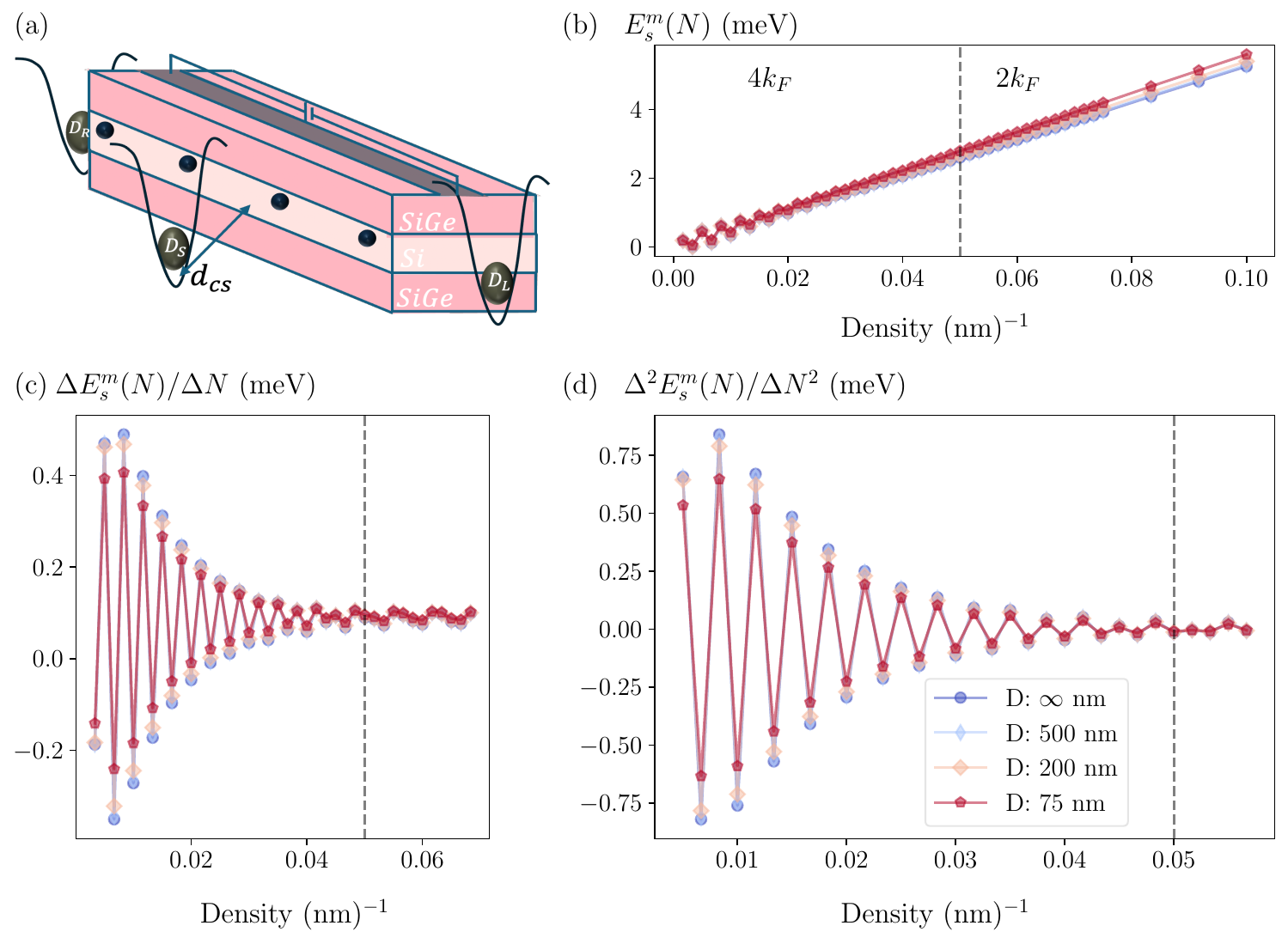}
    \caption{Charge sensing setup to probe the Wigner-Friedel crossover. (a) Setup of a charge sensing dot near the center of the interconnect that helps detect the Wigner-Friedel crossover with increasing density.
    $D_L(D_R)$ are quantum dots to the left (right) of the interconnect. (b) Energy of the ground state charge density measured by a single charge-sensor placed near the middle of the interconnect ($E_s^m$) for different screening lengths in the Si-channel. 
    The sensor measurements depict small oscillations at very low densities and then the energy measured follows a linear trend with the number of electrons. 
    The black vertical line at $n \approx 0.05 \text{ nm}^{-1}$ is the approximate density after which the system displays a Friedel-like $2k_F$ oscillation. (b) The first order difference $\Delta E_s^m(N)= E_s^m(N+1) - E_s^m(N)$ displays strong oscillations before the crossover density and lowers down to zero after the crossover. 
    (c) The second order difference in the sensor energy $\Delta^2 E_s^m(N)= E_s^m(N+1) - 2E_s^m(N) + 2E_s^m(N-1)$ also displays oscillations at low density, which reduces to zero after the crossover density.
 }
    \label{fig:charge-sensing}
\end{figure}

A charge sensor dot near the channel can measure signatures of the spatial ordering in the Wigner regime.
The Wigner regime  has a distinct $4k_F$ correlation profile where $N$ electrons in the channel would show $N$ peaks.
In \cite{shapirImagingElectronicWigner2019}, the charge density of a carbon nanotube housing a Wigner crystal was imaged by moving a single charge-sensing nanotube across the Wigner crystal.

In our work, since the Hamiltonian describing the interconnect has a reflection symmetry about the center \eqref{eqn:lattice-hamiltonian}, the charge density has to be a local minimum or maximum at the center. 
In the Wigner regime, as electrons are added to the channel, the charge density in the center rapidly oscillates between a maximum and near zero (Fig.~\ref{fig:spinful-screening}(b)). 
Here, strong repulsion between electrons leads to a wide separation, irrespective of spin. 
With increasing electron density, electrons are pushed closer to each other, and at a Wigner ratio $r_s \gtrsim 6$, electrons of opposite spin start pairing up.
As the system enters the Friedel-dominated $2k_F$ regime, there is accumulation of charges at the ends of the channel and the magnitude of the charge oscillations at the center reduce and saturate to a mean value.  

To capture these features and determine the crossover density, we propose a simple way to probe the electron charge density by placing a sensing quantum dot or a quantum point contact near the center of the interconnect.
Such a sensing quantum dot acts as a sensor of the local electric field and is sufficiently sensitive to measure the change in the local potential landscape due to the addition or subtraction of a single nearby charge.
When there is a high electron density near the middle of the interconnect, the shift in the Coulomb peak of the sensing dot is used as an estimate of the charge density in the interconnect.
We model the energy detected by a charge sensor placed a distance $d_{cs}$ from the center of the channel and a distance $d_{S}$ from the left end as:
\begin{align}
    E_s = \frac{e^2}{4\pi\epsilon_r \epsilon_0 d_{cs}}\int_0^L dx \frac{e^{-|x-d_{S}|/D_S} \quad n(x)}{\sqrt{1 + (x-d_{S})^2/d_{cs}^2}}
\end{align}
where $D_S$ is the effective screening length of the charge sensor.

Our results are presented in Fig.~\ref{fig:charge-sensing} where we choose $(d_{cs}, d_S, D_S): (100, 300, 25)$ nm.
Fig.~\ref{fig:charge-sensing}A depicts the setup and Fig.~\ref{fig:charge-sensing}B shows a linear increase in the energy detected by the sensor with increasing density in the channel.
In Fig.~\ref{fig:charge-sensing}C, in the low density Wigner regime, as the first few electrons enter the channel, we observe rapid oscillations in the measured energy with odd/even number of electrons in the channel as seen in Fig.~\ref{fig:charge-sensing}A.  
At densities $n \approx 0.05 \text{ nm}^{-1}$, near the Wigner-Friedel crossover density, the oscillations reduce and we observe flattening of $E^{m}_S$.  
The odd-even oscillation ($N \mod 2$) for energies in the Wigner regime gives way to an $N \mod 4$ oscillation in the spin-paired Friedel regime.
Gradually, as the channel enters the Friedel-dominated $2k_F$ regime ($n \geq 0.05 \;\mathrm{nm}^{-1}$), we observe the flattening of the first and second difference of $E^{m}_S$. 
As new electrons are added to the channel, no oscillations are observed in $E^{m}_S$ and the energy difference becomes closer and closer to the uniform density value $\approx 1/N$. 

\section{Channel with random alloy disorder}
\label{sec:results-disordered}

\subsection{Random alloy disorder}

\begin{figure}[t]
    \centering
    \includegraphics[width=\linewidth]{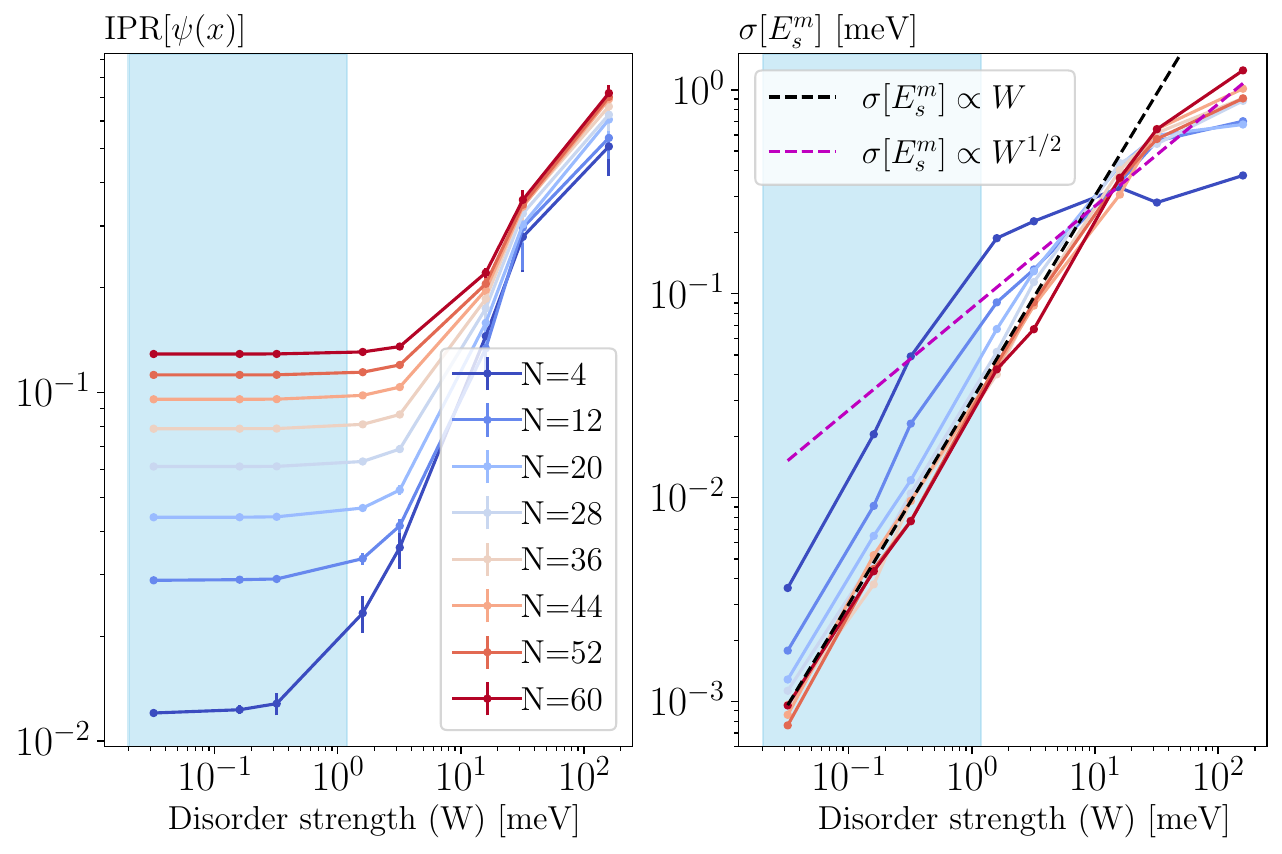}
    \caption{Short-range correlated disorder effects on the ground state in the channel. (a) Inverse participation ratio (IPR) of the ground state charge density for screening length D=75 nm with lattice disorder strength W. 
    As the strength of disorder increases, the IPR increases towards unity and the system undergoes Anderson localization. 
    Error bars represent the standard deviation over all disorder realizations.
    (b) Standard deviation (across realizations for each disorder strength) of the measured energy near the center of the channel for a charge-sensing experiment shows that
    a standard deviation greater than $\approx 5e^{-2} $ meV would obscure the suppression of the $N \text{ mod } 4$ oscillations in the Friedel regime, as seen in Fig~\ref{fig:charge-sensing}(c). 
    Blue shaded area is the current experimentally seen alloy disorder standard deviations for a $400\text{ nm}^2$ quantum dot ($W \approx (20 \text{ to } 400)$ \textmu eV), which scaled to our unit area is (80 to 1600) \textmu eV}
    \label{fig:random-aloy-disorder}
\end{figure}

In studying the effect of disorder, we first consider the random local potential of Eq.~\ref{eqn:short-range-disorder}.
To study the effect of disorder in the zero magnetic field limit, we compute the ground state for different disorder strengths $W$. For each realization, the initial state for the DMRG calculation is prepared by sequentially placing particles at the potential minima created by the disorder landscape and the electrostatic repulsion from previously placed particles. 
Our results for short-range disorder, averaged over 30 realizations for each $W$, are shown in Fig.~\ref{fig:random-aloy-disorder}.

We find that increasing disorder strength increasingly localizes the ground-state wavefunction. 
We characterize this via the inverse participation ratio (IPR) of the real-space charge density for spinful fermions
\begin{align}
    \text{IPR}[\psi(x)] = \frac{\int dx|\psi(x)|^4}{2 \int dx|\psi(x)|^2}
\end{align}
At low disorder, the Fourier transform of the real-space charge density, $n(k)$, has a distinct peak at $4k_F$ ($2k_F$) in the Wigner (Friedel) regime (per Fig.~\ref{fig:spinful-screening}(c,d,e,f)).
For disorder strengths approaching $\approx 500$ \textmu eV, additional random peaks emerge in $n(k)$, and the system eventually localizes (See Appendix \ref{sec:app-add-numerics}, Fig.~\ref{fig:onsite-disorder-crossover}).
To compare our disorder potential magnitude to those of seen in experiments, we relate the lattice-site disorder potential to the disorder potential for a notional quantum dot with area $\approx (20\text{ nm})^2$.
Such dots have disorder potential $\approx (20 \text{ to } 400)\;\text{}$\textmu ev~\cite{losert2023practical,Mcjunkin2021, mcjunkin2022sige, degli2024low}; 
the corresponding lattice-site disorder $W$ is $(80 \text{ to }1600)$ \textmu eV (see Appendix.~\ref{sec:device-parameters});

Our results are depicted in Fig.~\ref{fig:random-aloy-disorder}.
We find that the energy measured by a charge sensor near the center of the channel provides a robust experimental measure for disorder-induced localization. 
The standard deviation of this energy across disorder realizations increases sharply with $W$, following a power-law dependence before saturating upon full localization (Fig.~\ref{fig:random-aloy-disorder}b). 
This variance offers a method to estimate the disorder strength in a device, indicating that the Wigner regime melts for disorder strengths around $500$ \textmu eV. 
Systems in this localized regime may also be studied via Coulomb blockade experiments with an additional quantum channels~\cite{steinbergLocalizationTransitionBallistic2006}.
Having investigated alloy disorder, we now consider the effects of valley disorder.

\subsection{Valley splitting disorder}
In silicon heterostructures interconnects the sixfold valley degeneracy of bulk silicon is partially lifted by quantum confinement and strain, but a nearly degenerate excited valley state often remains accessible~\cite{BurkardEtAl2023}.
The energy splitting $E_v$ between these two low-lying states is not uniform; it varies significantly across the device and even between devices fabricated on the same chip. 
This disorder originates from the sensitivity of $E_v$ to the atomic-scale details of the heterostructure, particularly the Si-SiGe interface width and random alloy fluctuations.
As a result, the average splitting, $E_v$, can range widely from $10$ \textmu eV to over $ 500$ \textmu eV \cite{Goswami2007, losert2023practical,Mcjunkin2021, mcjunkin2022sige, degli2024low, Marcks2025-VSC, Volmer2024a}.
The distribution of these splittings is also affected by local fluctuations: sharp interfaces tend to produce larger, more uniform splittings, while wider interfaces lead to broader distributions centered at lower energies~\cite{losert2023practical}.
This valley splitting disorder is a critical parameter that can limit shuttling fidelities and, for our present interest, the Wigner-Friedel crossover.

In this section, we address the following question: how does this spatially-varying valley splitting disorder affect the charge-ordering of electrons, specifically the Wigner-Friedel crossover? We also propose experiments to probe if the Wigner-Friedel crossover can survive currently observed valley splitting disorder.

To investigate this, we consider the system in the high magnetic field limit, which effectively freezes the spin degree of freedom.
This reduces the cost of the DMRG and allows us to isolate the interplay between charge ordering in the Wigner regime and the valley degree of freedom. 
We leave the effects of spin-orbit coupling and valley-spin coupling to future work.

\begin{figure}[t]
    \centering
    \includegraphics[width=\linewidth]{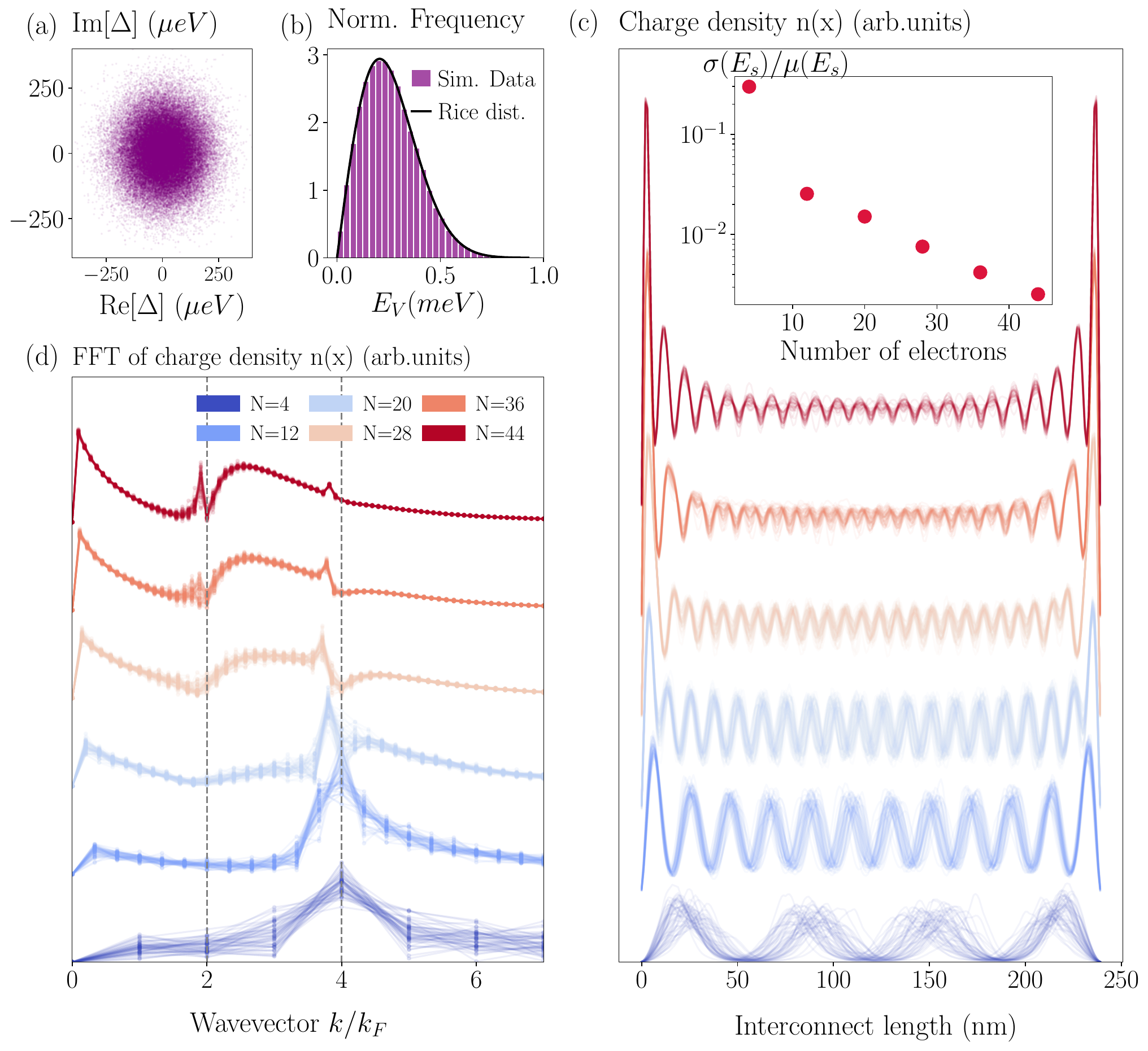}
    \caption{Effects of valley splitting disorder on the Wigner-Friedel crossover. (a) Distribution of the simulated intervalley coupling matrix element for a Si-SiGe device and (b) the distribution of the valley splitting (with total area normalized to unity) $E_v$, with the predicted Rice distribution $R(|\nu|, \sigma)$ with $\nu =  19.16$ \textmu eV and $\sigma=205$ \textmu eV, which we use in our simulations.
    (c) Ground state charge density for different electron densities in the interconnect for all valley disorder realizations reveals a crossover from $N$ peaks for $N$ electrons at low density (Wigner regime) to $N/2$ peaks at higher density (Friedel regime) where electrons of the same spin occupy the excited the valley state (charge density for each electron number is offset by an arbitrary constant). (Inset) The coefficient of variation ($\sigma/\mu$) of the energy detected by a charge sensor placed near the center of the channel for different number of electrons in the channel (each point is averaged over fifty disorder realizations) reveals the standard deviation is less than $1 \text{ }\%$ in for $N>20$ electrons (in the Friedel regime) (d) The crossover from the Wigner to Friedel regime visualized through the Fourier transform of the charge density, $n(k)$. 
    For each particle number in the interconnect, we plot the absolute value of the fourier transform of the charge density $|n(k)|$ for every realization.
    The maxima of the $|n(k)|$ occurs at $4k_F$ for $N \lesssim 24$, changing to near $2k_F$ at higher densities.}
    \label{fig:valley-disorder}
\end{figure}

Our results are summarized in Fig.~\ref{fig:valley-disorder}. 
Panels (a, b) illustrate the valley splitting disorder profile that we consider and its corresponding distribution showing a well-known Rice distribution, i.e. the distribution of magnitudes for a symmetric, bi-variate Gaussian distribution.
As in the previous section, the variance of the valley splitting energies on each lattice site in our simulations depends on the lattice cutoff; we detail the calculation of the variance in App.~\ref{appendix:valley-variance}.
The key results are in panels (c-d), which show the ground state charge density for 50 disorder realizations for different number of electrons in the channel.
At low densities, the system still clearly displays the characteristic $4k_F$ peak in $n(k)$, as seen in Fig.~\ref{fig:valley-disorder}d. 
We observe charge-density oscillations characteristic of the Wigner regime, even in the presence of strong valley disorder. As density increases, the system transitions to the Friedel regime.
The inset in Fig.~\ref{fig:valley-disorder}(c) further supports experimental validation.
We plot the coefficient of variation ($C=\sigma/\mu$) of the energy detected by a charge sensor (same as Sec.~\ref{sec:experiments}.B) where $\sigma$ is the standard deviation and $\mu$ is the average across disorder realizations.
The standard deviation of the energy measured by the charge sensor is much smaller than the mean and decreases to less than $1\%$ as the number of electrons goes beyond 20.
This indicates that the collective Wigner regime state is stable, and its macroscopic charge properties are not significantly disrupted by the expected levels of microscopic valley disorder.

Our results demonstrate that the Wigner regime at these densities survives current experimentally observed valley splitting disorder levels.
While we have focused on the spin-polarized case, this suggests that in the presence of an additional spin degree of freedom, a series of transitions (e.g., $8k_F \rightarrow 4k_F \rightarrow 2k_F$) could be observed as increasing density populates the spin and valley sectors. 
We thus propose that experiments can probe this robust Wigner regime state even in Si devices with valley splitting disorder.

\section{Capacitive coupling across the Interconnect}
\label{sec:cap-coupling-IC}
\begin{figure*}
\centering
    \includegraphics[width=\linewidth]{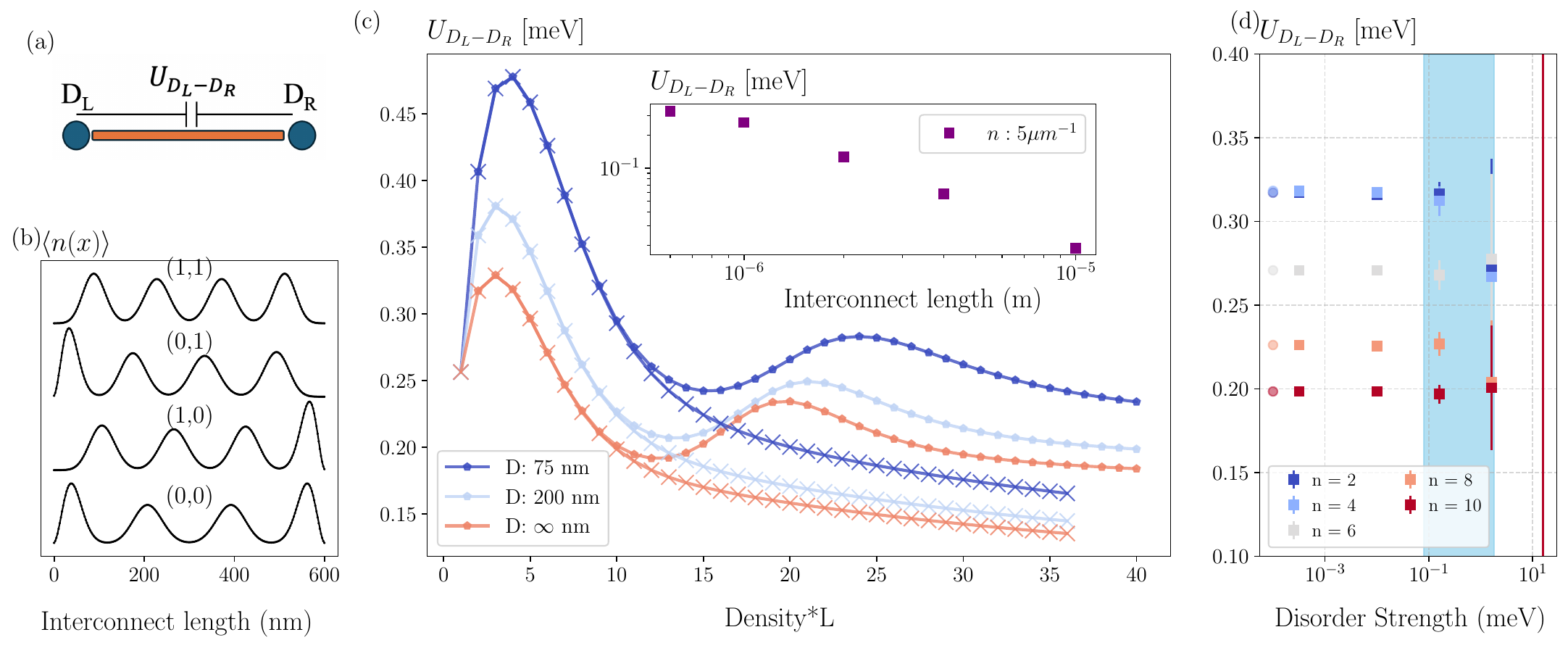}
    \caption{Capacitive energy ($U_{D_L-D_R}$) between quantum dots across the interconnect (a) depicted as a cartoon. 
    (b) Ground state charge density as a function of the filling of the quantum dots on the (left, right) of the channel.
    (c) Capacitive energy as a function of the filling of the interconnect for net zero (high) magnetic field in the interconnect depicted by diamond (cross) markers respectively. 
    As the electron density in the channel is increased, we observe two peaks in $U_{D_L-D_R}$ in the zero magnetic field limit. 
    In the low density regime, we observe a peak in the energy irrespective of the magnetic field, indicating a purely charge-sector phenomena. In this Wigner regime, Coulomb interaction with
    electrons in both the dots results in the ground state charge density being compressed as shown in panel (b). 
    The capacitive energy $U_{D_L-D_R}$ decreases with increasing density, as additional electrons screen the left charged dot from the right quantum dot. 
    In the zero magnetic field limit, as density increases, we observe a local maxima in the capacitive energy near the crossover density. 
    (Inset) The capacitive energy between quantum dots for different interconnect lengths shows an almost power-law decrease with increasing interconnect lengths.
    (d) The capacitive energy in the high magnetic field regime evaluated for different levels of onsite disorder. 
    Circles denote clean values, and squares denote values averaged over 50 disorder realizations. 
    Error bars are the standard deviation in the estimation of the mean ($\sigma/\sqrt{N}$).
    We observe that the capacitive coupling survives disorder levels up to $W \leq 1 $ meV where the blue patch denotes currently experimentally observed disorder levels ($80-1600$) \textmu eV.}
    \label{fig:cap_coupling}
\end{figure*}

Electrons in the channel couple electrostatically to dots at either end of the channel.
This leads to an effective capacitive coupling \cite{Chan2002,Srinivasa2015} between dots $L,R$ near the left and right ends of the channel:
when an electron enters dot $L$ the charge density in the channel shifts towards dot $R$, increasing the energy of an electron in that dot.
Such long-range capacitive coupling between quantum dots can be leveraged to execute high-fidelity two-qubit gates between qubits with a non-vanishing electric dipole, like the singlet-triplet, flopping mode, or exchange-only qubits~\cite{Trifunovic2012, Srinivasa2015, Cayao2020, Feng2021}.  

To estimate the capacitive energy between electrons within quantum dots on either side of the channel, we introduce an additional term in the interconnect Hamiltonian \eqref{eqn:lattice-hamiltonian}, given by:
\begin{flalign}
    \label{eqn:cap-coupling-eq1}
    H_{D^{\alpha}-IC} &= \sum_{i,\alpha} V(x_{D^{\alpha}}, x_i) \hat{n}_{D^{\alpha}} \hat{n}_i
    \\
    V(x_{D^{\alpha}}, x_i) &= \frac{e^2}{4\pi \epsilon |x_{D^{\alpha}} - x_i|}
\end{flalign}
where $D^\alpha= \{L, R\}$ represents the left or the right quantum dot. 
We take each quantum dot to be 25nm from the ends of the interconnect.

To understand the effect of the interconnect on neighboring quantum dots, we first calculate the capacitive coupling $U_{D_L-D_R}$ between the dots due to the channel as a function of the density in the interconnect. 
To do so we compare the ground state energies of the interconnect when the configuration of the left (right) dots are (0,0), (1,0), (0,1) and (1,1). 
The coupling is then
\begin{equation}
    \label{eqn:cap-coupling}
    U_{D_L-D_R} = U^{1,1}_g - U^{1,0}_g - U^{0,1}_g + U^{0,0}_g
\end{equation}
where $U^{a,b}_g$ is the ground state of the system with $a(b)$ electrons in the left (right) quantum dot.
If the two dots were completely independent, then
the energy change due to adding an electron in the left and right dots would be additive, so the coupling $U_{D_L-D_R}$ would be zero.
We calculate how $U_{D_L-D_R}$ changes with the density of electrons in the interconnect, across the Wigner-Friedel crossover, in zero and high external magnetic fields.

Our results are shown in Fig.~\ref{fig:cap_coupling}. 
In the spinful (zero magnetic field) case, the capacitive coupling $U_{D_L-D_R}$ shows a distinctive two-peak structure as a function of electron density.
The first peak occurs consistently near \( nL \approx 3{-}4 \), deep in the Wigner regime.
In this low-density limit, electrons are spatially ordered.
The second occurs near the maximum of compressibility (i.e. the minimum of the inverse compressibility shown in Fig.~\ref{fig:compressiblity}(A)), and close to the Wigner-Friedel crossover point.
Beyond the second peak, the many-body ground state of the interconnect becomes a dense Luttinger liquid that screens interactions between the dots.
In addition, we see that that $U_{D_L-D_R}$ is the same for both high and zero magnetic fields in the interconnect.

In addition to the clean system, we also investigate the effects of short-ranged correlated disorder, as described in \eqref{eqn:short-range-disorder}, on the capacitive coupling between dots. 
We evaluate the capacitive coupling for each of 50 disorder realizations; 
the results of which are depicted in Fig.~\ref{fig:cap_coupling}(d).
In the high magnetic field limit, we see that the capacitive coupling is robust until $W \approx 1$ meV and then starts to decrease.
Furthermore, we also see that the capacitive coupling remains robust at currently observed experimental disorder levels.
At disorder strengths of $\approx 10^1$ meV, the device is essentially disorder-dominated, and $U_{D_L-D_R}$ shows strong deviations from the clean values. 

Our results on the capacitive coupling between quantum dots across the interconnect describes another potential experiment to probe the Wigner-Friedel crossover. 
Our results on capacitive coupling in the Wigner regime can be experimentally probed through the charge stability map of the left- and right- quantum dots. 
If a double quantum dot charge stability map is formed using the left and right quantum dots, then the inclination of the charge transition line informs the capacitive coupling between the dots. 
A high capacitive coupling that is mediated by the Wigner regime in the interconnect would result in high inclination/slope of the charge transitions in the charge stability map whereas a low coupling (Friedel regime) would correspond to the transition lines being almost orthogonal. 
Moreover, operating the interconnect in the Wigner regime with a high capacitive energy ($200 \text{ to }450$) \textmu eV can be beneficial for executing long-range capacitively coupled two qubit gates between qubits across an interconnect. 
Finally, we note that the capacitive coupling values that we obtain are comparable to those obtained from superconducting resonators~\cite{Burkard2020}. 

\section{Summary and Discussion}
\label{sec:summary}

Our results show that silicon interconnects are suitable platforms to observe many-body physics in the Coulomb-energy dominated Wigner regime. 
We have shown that electrons in a one-dimensional channel formed within a Si-well, interacting via the screened Coulomb interaction, display strongly interacting mesoscopic physics, particularly the Wigner regime at low-enough densities and disorder strengths. 

First, we considered a clean interconnect with no disorder. 
Through density matrix renormalization group (DMRG) calculations, we showed that at low electron densities $n \lesssim a_B^{-1}$ the ground state of the interconnect displays strong $4k_F$ oscillations with strong electronic compressibility. 
As the density is increased through $n \approx a_B^{-1}$, we observed a crossover from the Wigner regime to the kinetic energy-dominated Friedel regime, in which the largest oscillations are at $2k_F$.

Second, to probe the crossover, we proposed charge transport and charge sensing experiments. 
We supplemented our numerical results on electronic compressibility with results from bosonization which indicate similar behavior with increasing density.
The electronic compressibility $\kappa$ of the interconnect, defined as the inverse second-order difference of the ground state energies with respect to the number of electrons, $\kappa^{-1}(n) \propto \partial^2 E / \partial n^2$, can be probed by measuring conductance through the interconnect as the interconnect gate potential is varied between the source and drain potentials.
The spacing between the conductance peak gate voltages, $\Delta V_G$ obtained in a Coulomb blockade experiment is proportional to the incompressibility of the many-body state formed within the interconnect $\Delta V_G = \kappa^{-1}(N)$. 
In the low-density Wigner regime (high $r_s$), we find that the inverse compressibility  is small, leading to a small spacing between conductance peaks.
In the high-density Friedel regime (low $r_s$), $\kappa^{-1}$ is large, leading to a large spacing between conductance peaks. 
This dependence of $\kappa^{-1}$ on electron density $n$—increasing with density across both the Wigner and Friedel regimes—serves as a key experimental signature for identifying the formation of a Wigner crystal.
To further determine the crossover density, we proposed charge-sensing experiments with a charge-sensor placed near the center of the interconnect. 
Since the system has reflection symmetry about the center, the ground state charge density about the center of the channel is either a minimum or a maximum. 
This is more clear in the Wigner regime where oscillations with odd (even) electron fillings in the interconnect occur, which subsequently reduce and vanish as the system enters the Friedel regime. 
In the Wigner regime ($4k_F$ correlations), the oscillations occur with addition of every electron, whereas in the Friedel regime, the oscillation with particle number vanishes.

Third, we evaluated the Wigner-Friedel crossover in the presence of short-range-correlated disorder, which heuristically models local defects and random alloy disorder at the Si-SiGe interface.
We first considered the disorder to be a random local chemical potential;
we then considered coupling to a valley degree of freedom.
In both cases we found that the crossover was unaffected by disorder up to $W \approx 500$ \textmu eV typically seen in experiments with dot sizes of $r\approx 20\text{ nm}$;
for stronger disorder, the disorder potential dominates and the ground state charge density is best understood as localized.

Fourth, we proposed that dense electrons in the channel can capacitively couple electrons in quantum dots on the ends of the channel. 
We showed that the Wigner regime in the channel increases the capacitive coupling between quantum dots on either side of the channel-- the occupancy of the left dot enhances the electrostatic energy of the right dot. 
The poor screening capability of the crystal-like regime allows a charge perturbation on a source dot to strongly influence the electrostatic environment of a distant target dot, an effect that diminishes as the system enters the denser, highly-screened Friedel regime.
We found that the capacitive energy between the dots increases to a maximum in the low-density Wigner regime and then decreases with increasing density.
In addition, we evaluated the capacitive coupling in the presence of short-ranged correlated disorder, and found that the coupling is robust to disorder levels up to $1000$ \textmu eV and before reducing towards zero.  
The Wigner regime in the interconnect thus provides a novel way to tune and couple quantum dots without resorting to exchange interactions.

We have focused on the ground state of a clean system with short-range correlated disorder for a medium range interconnect of length $L \geq 600 \text{ nm}$.
Future theoretical work should incorporate a more detailed treatment of effects neglected herein, such as spin-orbit coupling, valley-spin coupling, and extrinsic charge noise.
Further investigation into the device's geometric parameter space is also warranted. 
Longer interconnects are expected to further reduce electron density and enhance Coulomb repulsion, thereby strengthening the Wigner regime. 
Conversely, wider channels may reduce confinement and enable the study of more exotic phases, such as zig-zag Wigner crystal regimes. 
In the future, more realistic atomistic studies that include long-range correlated charge disorder, and dynamical disorder effects could also be useful. 

In the context of the suggested experimental probes, we discuss common effects that could alter our results. 
We have not evaluated the effect of disorder on the spacing between conductance peaks, and disorder can modify the peak spacing, such that the increasing spacing between conductance peaks may not be clearly visible.
Furthermore, the charge sensing method used to detect the crossover density can be sensitive to the sensor's proximity and its effective screening length; its utility reduces as the distance of the sensor from the channel exceeds its screening length.
The capacitive energy between quantum dots across the interconnect would also depend on the exact distance between dots and the interconnect, and the optimal device geometry for a specific experiment will need to be further evaluated. 
Future work can focus on understanding the entanglement dynamics of spin qubits due to the Wigner regime mediated capacitive coupling and its benefits over traditional schemes.
In addition, the long-range capacitive coupling between electrons mediated by the Wigner regime can be probed in platforms such as electrons on liquid helium~\cite{Koolstra2019-lk}, or carbon nanotubes~\cite{deshpandeElectronLiquidsSolids2010}.

Finally, it is important to understand the range of electron densities required to observe the Wigner regime and determine if such densities are achievable in current-era silicon devices.
In this work, we find that the Wigner regime is present up to densities of $10^{11} \text{ cm}^{-2}$, which is substantially higher than the percolation transition density reported for Si-SiGe devices~\cite{degli2024low}.
This order-of-magnitude margin suggests that the crossover is experimentally accessible, even if the required densities are on the lower end of the current spectrum.
With rapid advancements in device fabrication, our results highlight that silicon interconnects not only offer a path toward scalable quantum computing but also serve as a robust and promising platform for exploring fundamental mesoscopic physics.

\begin{acknowledgments}
The authors are grateful to Merritt Losert, Srilekha Gandhari, Stuart Yi-Thomas, Jiayao Zhao, David Kanaar, Hongwen Jiang, Joseph Salfi, and Vanita Srinivasa for fruitful discussions.
A.S.R acknowledges the University of Maryland supercomputing resources (https://hpcc.umd.edu) made available for conducting the research reported in this paper.
This research was sponsored in part by the U.S. Army Research Office
(ARO) grant W911NF-23-1-0242 and W911NF-
23-10258 and NSF QLCI award OMA-2120757.  This work was performed in part at the Kavli Institute for Theoretical
Physics (KITP), which is supported by grant NSF PHY-2309135.
This work was completed while C.D.W. held an NRC Research Associateship award at the United States Naval Research Laboratory. S.R.M. is supported by the NSF QLCI (award No. OMA-2120757).Technology.

\end{acknowledgments}

\appendix

\section{Parameters, units and non-dimensionalization}
\label{sec:device-parameters}

We seek to model a continuum Hamiltonian ($H_C$) for electrons in the silicon interconnect with a discrete lattice Hamiltonian ($H_L$) given by : 
\begin{align}
    H_c &= -\sum_i^N \frac{\hbar^2 k_i^2}{2m_t} + \frac{e^2}{4\pi \epsilon_r \epsilon_0 d}\sum_{j,k>j} \Gamma(x_j, x_k) \\
    H_L &= -t \sum_i^N (\hat{c}^{\dagger}_i\hat{c}_{i+1} + \hat{c}^{\dagger}_{i+1}\hat{c}_{i}) + A\sum_{j,k>j} \Gamma(x_j, x_k)
\end{align}
The natural energy units for the system can be obtained by comparing the spectrum of the lattice model (at low energies) to the continuum, i.e using $\hat{c}_j \approx e^{-ikja}$. we get $t \approx \hbar^2/(2m_ta^2)$ where `a' is the lattice constant. 
Using the Fourier transform to go from the real-space to momentum-space representation, we use : $\hat{c}_j = \frac{1}{\sqrt{N}} \sum_k \hat{c}_k e^{ikja}$ and $\hat{c}_j^{\dagger} = \frac{1}{\sqrt{N}} \sum_k \hat{c}_k^{\dagger} e^{-ikja}$ in $H_L$, we get, 
\begin{align}
\begin{split}
    H_L(KE) &= -t \sum_j \left( c^{\dagger}_j c_{j+1} + c^{\dagger} _{j+1} c_{j}\right)\\
    &= - 2t \sum_k \cos{(ka)} c^{\dagger}_k c_k
    \\
    &= - 2t \left(1 + \frac{(ka)^2}{2}\right) \hat{n}_k \text{  (keeping low energy terms)} \\
    &\approx - ta^2 k^2  \hat{n}_k\text{ (constant can be ignored)} \\
    &= - \hbar^2k^2/2m_t \hat{n}_k
\end{split}
\end{align}
Using this, we see that the hopping term (t) is set by the effective transverse mass of the electron in silicon $(m_t: 0.19m_e)$ and the lattice spacing $(a)$ : $t = \hbar^2/2m_ta^2$. 
These fundamental units of length and energy in our simulations are given in Table\eqref{tab:my_table}.

To observe the Wigner-Fridel crossover for densities tractable within DMRG, we consider a channel of length $L=600$ nm. 
To ensure that the channel length, L is much larger than the width $d$, we take $L/d \gg 10$.
The other derived units that we use in our simulations are given in Table\eqref{tab:my_table_2}.
From our analysis, we observe the Wigner-Friedel crossover at at density $n\approx 5 \times10^{11} \text{ cm}^{-2}$ for number of electrons $N \approx 30$. 
For interconnects of length $L \approx 1$ \textmu m, and width $d \approx 100 \text{ nm}$, the crossover would occur at $N \approx 400$.
The unit area, $A$ that we consider for our disorder models is $A=a \times d = 25 \text{ nm}^2$
\begin{table}[h]
    \centering
    \begin{tabular}{|lcr|}
    \hline
    Lattice spacing & a & 2.5 nm \\
    Unit of energy (t) & $\frac{\hbar^2}{2m_ta^2}$ & $32.084$ meV \\
    \hline
    \end{tabular}
    \caption{Fundamental units for our numerical calculations}
    \label{tab:my_table}
\end{table}

\begin{table}[]
    \centering
    \begin{tabular}{|lcccr|}
    \hline
    Quantity & Symbol & Value & SI & Non-Dim \\
    \hline
    Lattice length & L & L/a& 600 nm  & 240\\
    Channel width & d & d/a& 10 nm  & 4\\
    Coulomb strength & A & $\frac{e^2}{4\pi\epsilon_r\epsilon_0 d}$& 12.52 meV& 0.38 \\
    Rel. permittivity of Si & $\epsilon_r$ & 11.7 & 11.7 & 11.7\\
    \hline
    \end{tabular}
    \caption{Derived units for our numerical calculations}
    \label{tab:my_table_2}
\end{table}

\section{Asymptotic behavior of the interaction potential}
\label{appendix:bosonization}

Here, we study approximate expressions for the Fourier transform of the interaction potential, focusing on logarithmic divergences arising UV and IR cutoffs in the system, relevant for our studies of the inverse compressibility. In the main text, we simply use the numerically computed discrete Fourier transform, 
\begin{align}
    \tilde{U}(q) = \frac{1}{N_{tot}}\sum_{i,j} U(x_i - x_j) \cos(q(x_i - x_j)).
\end{align}
for our interaction potential $U(r)$ in Eq.~\eqref{eqn:screened-interaction}.
However, to gain analytic insight into the behavior of this function, here we approximately compute the Fourier transform as
\begin{align}
    \tilde{U}(q) \simeq 2\int_0^{L/2} dr \cos(qr) U(r),
\end{align}
where we impose an IR cutoff $L$ arising from the finite extent of the system. 

In the limit where $L \gg d$, and using \eqref{eqn:screened-interaction} for $U(r)$, we then have the approximation
\begin{align}
    \tilde{U}(q) \simeq 2V_0\, \mathfrak{R}\Bigg\{\frac{\pi}{2} \left[H_0\left(\tilde{z}\right) - Y_0\left(\tilde{z}\right) \right] - \Gamma\left(0,\tilde{L}\tilde{z}/2\right)\Bigg\},
\end{align}
where $\mathfrak{R}$ denotes the real part and $\tilde{z} \equiv 1/\tilde{D} + i\tilde{q}$, and we have defined dimensionless variables $\tilde{D} \equiv D/d$, $\tilde{q} \equiv qd$, and $\tilde{L} \equiv L/d$. Here, $H_0$ is the Struve function, $Y_0$ is the Bessel function of the second kind, and $\Gamma$ is the incomplete Gamma function.

To gain insight into the behavior at low densities, we consider the large screening limit $D \gg L$, and are also interested in the case where $\tilde{q} \ll 1$, relevant for considering $q=0$ or $q = 2k_F$ in the low-density limit. Under these assumptions, we then have the asymptotic expansions 
\begin{gather}
    \mathfrak{R}\left\{\frac{\pi}{2}\left[H_0(\tilde{z}) - Y_0(\tilde{z})\right]\right\} \sim -\log\left(|\tilde{z}|/2\right) - \gamma\\
    \Gamma(0,\tilde{L}\tilde{z}/2) \sim \frac{2}{\tilde{L}\tilde{z}}e^{-\tilde{L}\tilde{z}/2}
\end{gather}
up to $O(\tilde{z})$. In practice, we find these expressions also provide a decent qualitative approximation for $D \gtrsim L$ as the system becomes unscreened. Thus, we obtain the following expressions in the limit $n\rightarrow 0$:
\begin{gather}
    \tilde{U}(0) \sim 2V_0 \left[\log(2\tilde{D}) - \gamma - \frac{2D}{L} e^{-L/2D}\right]\\
    \begin{split}
        \tilde{U}(2k_F) \sim 2V_0 \Big[\log\left(\frac{2\tilde{D}}{\sqrt{1 + (2k_F D)^2}}\right) - \gamma \\
        - \frac{2D}{L\sqrt{1 + (2k_F D)^2}} e^{-L/2D}\Big].
    \end{split}
\end{gather}
Thus, we observe that $\tilde{U}(2k_F)$ is weakly dependent on the density, and in the large screening limit, the first term provides a logarithmic divergence as $k_F \propto n\rightarrow 0$.

\section{MPS methods}

To identify the ground state attained by a set of spinful/spinless electrons in the one-dimensional channel, we model the Hamiltonian of the channel as a matrix product operator (MPO) and find the ground state as a matrix product state (MPS), by iteratively minimizing the ground state energy using the DMRG algorithm\cite{schollwoeckDensitymatrixRenormalizationGroup2011}. The Hamiltonian for the quasi 1D channel is given by \eqref{eqn:lattice-hamiltonian}. 

\begin{figure}[]
    \centering
    \includegraphics[width=\linewidth]{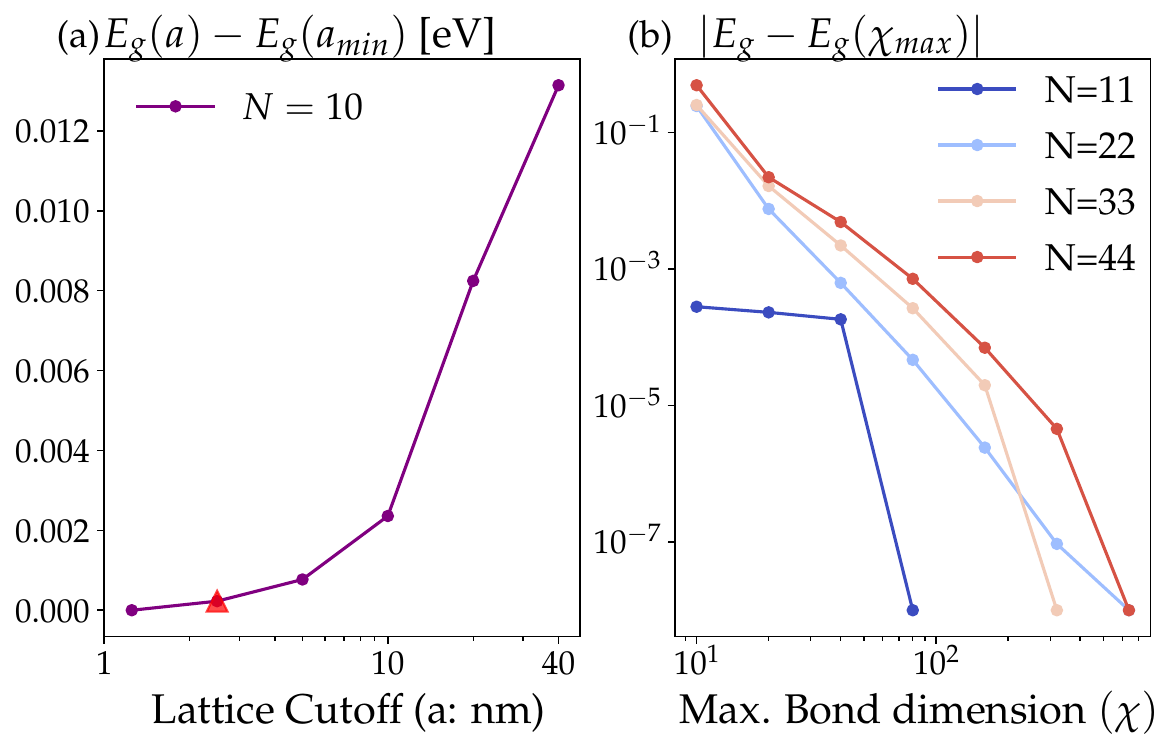}
    \caption{DMRG convergence results (a) Ground state energy of the interconnect from DMRG for different lattice cutoffs for the same number of particles ($N=10$) and length of interconnect ($L=600\text{ nm}$). Red triangle shows the lattice cutoff $2.5\text{ nm}$ we use in the main text. (b) Variation of the difference between ground state energies for different maximum bond dimensions shows that our simulations are sufficiently converged i.e $ |E_g(\chi) - E_g(\chi_{\text{max}})|< 1e-8$ for $\chi > 650$. So we use a maximum bond dimension of 650 in our DMRG calculations until ground state energy cutoff of $10^{-8}$ is satisfied.}
    \label{fig:bond-dim-convg}
\end{figure}

Having created the MPO, the ground state MPS is obtained from the DMRG algorithm using the ITensor package~\cite{fishmanITensorSoftwareLibrary2022}. 
The bond-dimension convergence and lattice cutoff convergence results are shown in Fig.\ref{fig:bond-dim-convg}.
For simulating a net zero magnetic field in the device, we look at the ground state in the spin sector $s_z= 0$ and for the limit of high magnetic field we consider the ground state of spinless fermions, which would correspond to polarized spinful electrons.

\section{Valley disorder calculations}
\label{appendix:valley-variance}

\begin{figure}
    \centering
    \includegraphics[width=\linewidth]{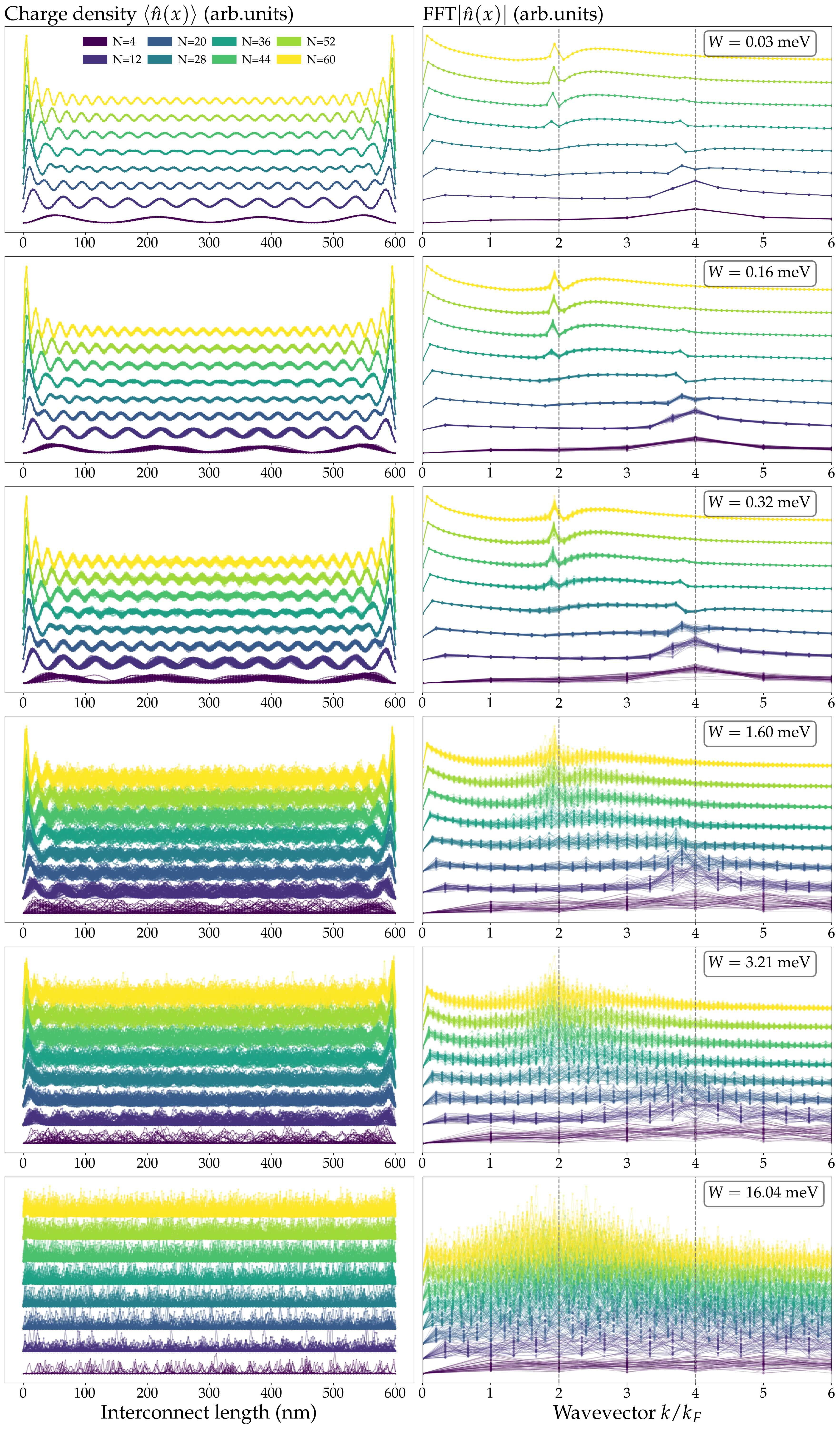}
    \caption{Wigner-Friedel crossover in the presence of random alloy disorder as background onsite potential.
    As seen in Sec.\ref{sec:results-disordered}.A, the Wigner regime with dominant wavevector at $4k_F$ crosses over to the Friedel regime with dominant correlations at $k=2k_F$.
    Each row in the above plot corresponds to a particular lattice disorder level with 30 disorder realizations for each particle number and for each onsite disorder level.
    For each particle number, the corresponding charge density and fourier transform are offset by a constant.
    At low lattice disorders ($W< 1$ meV), the crossover is clear and can be studied through charge sensing and transport. 
    At higher lattice disorder levels ($W > 1$ meV), disorder suppresses the crossover.
    When the system is disorder dominated, the ground state of the system localizes with IPR[$\psi(x)$] $\approx 1$.} 
    \label{fig:onsite-disorder-crossover}
\end{figure}

We can rewrite \eqref{eqn:valley-ham-2} using second-quantization operators.
By mapping the Pauli matrices to the occupation basis of the two valley minima (indexed by $\tau=1,2$), we define: 
\begin{align}
    \begin{split}
    \hat{\tau}^x_i &= \hat{c}^\dagger_{i,1}\hat{c}_{i,2} + \hat{c}^\dagger_{i,2}\hat{c}_{i,1}\\
    \hat{\tau}^y_i &= -i\hat{c}^\dagger_{i,1}\hat{c}_{i,2} + i\hat{c}^\dagger_{i,2}\hat{c}_{i,1}\\
    \hat{\tau}^z_i &= \hat{c}^\dagger_{i,1}\hat{c}_{i,1} - \hat{c}^\dagger_{i,2}\hat{c}_{i,2}
    \end{split}
\end{align}
Substituting these into the Hamiltonian in \eqref{eqn:valley-ham-2}, we get the more compact result:
$$\hat{H}_{valley} = \sum_{j=1}^{N_{sites}} |\Delta_{v}(x_j)| \left( e^{-i\phi(x_j)} \hat{c}^{\dagger}_{j,1} \hat{c}_{j,2} + e^{+i\phi(x_j)} \hat{c}^{\dagger}_{j,2} \hat{c}_{j,1} \right)$$
The valley Hamiltonian thus takes the form of a local mixing term where the complex coupling $\Gamma_j = |\Delta_v(x_j)|e^{-i\phi(x_j)}$ encodes both the magnitude of the splitting and the phase of the interface. This form highlights that valley disorder acts as an on-site scattering source, enabling an electron to `hop' from valley 2 to valley 1 (and vice versa) with a position-dependent phase.

We now describe our calculations for evaluating the inter-valley coupling matrix element. 
We follow the same approach as described in \cite{losert2023practical} and we estimate the mean and variance of valley splitting due to random alloy disorder. 
The variance is known to depend on the effective germanium profile along the z-axis and the unit area being averaged over.
The mean valley splitting ($\Delta_0$) and its variance are given by:
\begin{align}
    \begin{split}
    \Delta_0 &= \frac{a_0 \Delta E_c}{4 (X_w - X_s)}\sum_l e^{-2ik_0z_l} (\bar{X}_l - X_s) |\psi_{env}(z_l)|^2 \\
    \text{Var}(\Delta) &= \frac{a_0^2 \Delta E_c^2}{2N_{\text{eff}}(X_w - X_s)^2} \sum_l \bar{X}_l(1 - \bar{X}_l)|\psi_{env}(z_l)|^4 \\
    N_{\text{eff}} &= \frac{a \times d}{a_0^2}
    \end{split}
\end{align}
where $X_l$ is the silicon concentration in layer $l$ and $X_{w/s}$ is the silicon concentration in the well/substrate respectively, which we take to be 1/0.7,
$k_0 = 0.82 (2\pi/a_0)$ is location of the valley minima in the reciprocal space,
$a_0$ is the size of a unit cell in silicon, $a_0 = 0.542 \text{ nm}$,
$\psi_{env}(z_l)$ is the wavefunction in the silicon well, 
and $N_{\text{eff}}$ is the effective number of silicon atoms in our considered unit area with lattice constant $a$ and channel width $d$.
$\Delta E_c$ is the conduction band offset between silicon and germanium~\cite{losert2023practical}. 

For these calculations, we model the silicon quantum well as a well with a depth of $10$ nm and linear ramps of $1$ nm at the ends of the Silicon well. 
The barrier consists of 30\% germanium. 
We solve for the ground state in such a potential and find $\psi_{env}(z_l)$, and then evaluate $\Delta_0, \text{ and } \text{Var}(\Delta)$.
For an effective area $A = a\times d = 25 \text{ nm}^2$, we get $\Delta_0 = 9.58$ \textmu eV and $\sigma(\Delta) = 1.03$ meV. 
Since $\text{Var}(\Delta) \times A$ is a constant, for a quantum dot of area $\pi*400\text{ nm}^2$, the standard deviation in the valley disorder is $\approx 145$ \textmu eV. 
Similar disorder levels are seen in real quantum dots.

It is well known that the valley splitting follows a Rician distribution.
A distribution $R \approx \text{Rice}(|\nu|, \sigma)$, if $R=\sqrt{X^2 + Y^2}$ where $X \approx N(\nu \cos(\theta), \sigma^2)$ and $Y \approx N(\nu \sin(\theta), \sigma^2)$ are statistically independent normal random variables and $\theta$ is any real number (for our simulations, we choose $\theta=2^{-1/2}$).
This means that we can approximate the intervalley coupling, $\Delta$ as a circular Gaussian random variable centered at $\Delta_0$, for which the variances of the real and imaginary parts of are both (1/2)Var[$\Delta$].
Thus, we relate the derived mean and variance in valley disorder $(\Delta_0, \text{Var}(\Delta))$ to the Rician distribution (the fitted rice curve is shown in Fig.~\ref{fig:valley-disorder}(B)), $R(|\nu|, \sigma)$ as $|\nu| = 2\Delta_0$ and $\sigma = \sqrt{2}\text{Var}(\Delta)$. 
Since this variance changes with the effective area averaged over, for a typical quantum dot with radius $20 \text{ nm}^2$, the corresponding Rice distribution parameters will be $\nu = 20 $\textmu eV and $\sigma= 205.32$ \textmu eV.

\section{Additional Numerical results}
\label{sec:app-add-numerics}

In Fig.~\ref{fig:onsite-disorder-crossover}, we discuss additional results on the effects of random alloy disorder on the Wigner-Friedel crossover. 
The crossover can be probed through the standard deviation in the charge sensor enegy as seen in Fig.~\ref{fig:random-aloy-disorder}(B). 
In Fig.~\ref{fig:onsite-disorder-crossover}, we show how the dominant wavevector changes from $4k_F \rightarrow 2k_F$ with increasing electron density in the channel for increasing disorder levels. 
Our lattice disorder model, $W$ is the average background potential for a unit area of $A = 25 \text{ nm}^2$ that we use in our simulations.
To compare $W$ with $W_{QD}$, the alloy disorder levels seen in a typical Si-SiGe quantum dot of area $A_{QD} = 400 \text{ nm}^2$, we use $W_{QD}\times \sqrt{A_{QD}} \approx W \times \sqrt{A_l}$, obtaining $W_{QD} = W/4$. 
From these calculations, the Wigner-Fridel crossover must survive until experimentally observed disorder levels of $W_{QD} \leq  800$ \textmu eV.

\bibliography{references}

\end{document}